\documentclass[aps,pra,twocolumn,reprint,superscriptaddress,nofootinbib,amsmath,amssymb,amsfonts,nolongbibliography]{revtex4-2}
\usepackage{amssymb}
\usepackage{amsmath}
\usepackage{amsthm}
\usepackage{bm}
\usepackage[svgnames]{xcolor}
\usepackage{graphicx}
\usepackage{physics}
\usepackage{float}
\usepackage{tabularx}
\usepackage{dcolumn}
\usepackage{isotope}
\usepackage{color}
\usepackage{bbold}
\usepackage{lipsum}

\usepackage{soul}
\colorlet{MAGENTA}{magenta}

\providecommand{\ket}[1]{|#1\rangle}

\newcommand{\ua}{\uparrow}
\newcommand{\da}{\downarrow}
\newcommand{\mass}{\mathfrak{m}}
\usepackage[ruled,vlined]{algorithm2e}
\usepackage{algorithmicx}

\newcommand{\abst}[1]{\ensuremath{| #1 |}}
\theoremstyle{definition}
\newtheorem{res}{Result}
\newtheorem{rem}{Remark}

\usepackage[hyperfootnotes=false,breaklinks=true]{hyperref}

\begin{document}

\title{Optimizing two-qubit gates for ultracold atoms using Fermi-Hubbard models}

\author{Juhi~Singh\,\href{https://orcid.org/0000-0001-9807-9551}{\includegraphics[height=6pt]{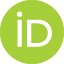}}}
\email[]{j.singh@fz-juelich.de}
\affiliation{Forschungszentrum Jülich GmbH, Peter Grünberg Institute, Quantum Control (PGI-8), 52425 Jülich, Germany}
\affiliation{Institute for Theoretical Physics, University of Cologne, 50937 Köln, Germany}

\author{Jan~A.~P.~Reuter\,\href{https://orcid.org/0009-0008-6786-4053}{\includegraphics[height=6pt]{ORCID-iD_icon-64x64.png}}}
\affiliation{Forschungszentrum Jülich GmbH, Peter Grünberg Institute, Quantum Control (PGI-8), 52425 Jülich, Germany}
\affiliation{Institute for Theoretical Physics, University of Cologne, 50937 Köln, Germany}

\author{Tommaso~Calarco\,\href{https://orcid.org/0000-0001-5364-7316}{\includegraphics[height=6pt]{ORCID-iD_icon-64x64.png}}}
\affiliation{Forschungszentrum Jülich GmbH, Peter Grünberg Institute, Quantum Control (PGI-8), 52425 Jülich, Germany}
\affiliation{Institute for Theoretical Physics, University of Cologne, 50937 Köln, Germany}
\affiliation{Dipartimento di Fisica e Astronomia, Università di Bologna, 40127 Bologna, Italy}

\author{Felix~Motzoi\,\href{https://orcid.org/0000-0003-4756-5976}{\includegraphics[height=6pt]{ORCID-iD_icon-64x64.png}}}
\affiliation{Forschungszentrum Jülich GmbH, Peter Grünberg Institute, Quantum Control (PGI-8), 52425 Jülich, Germany}
\affiliation{Institute for Theoretical Physics, University of Cologne, 50937 Köln, Germany}
\author{Robert~Zeier\,\href{https://orcid.org/0000-0002-2929-612X}{\includegraphics[height=6pt]{ORCID-iD_icon-64x64.png}}}
\email[]{r.zeier@fz-juelich.de}
\affiliation{Forschungszentrum Jülich GmbH, Peter Grünberg Institute, Quantum Control (PGI-8), 52425 Jülich, Germany}

%\date{\today}
\date{June 17, 2025}

\begin{abstract}
Ultracold atoms trapped in optical lattices have emerged as a scalable and promising platform for quantum simulation and computation. However, gate speeds remain a significant limitation for practical applications. In this work, we employ quantum optimal control to design fast, collision-based two-qubit gates within a superlattice based on a Fermi-Hubbard description, reaching errors in the range of $10^{-3}$ for realistic parameters. Numerically optimizing the lattice depths and the scattering length, we effectively manipulate hopping and interaction strengths intrinsic to the Fermi-Hubbard model. Our results provide five times shorter gate durations by allowing for higher energy bands in the optimization, suggesting that standard modeling with a two-band Fermi-Hubbard model is insufficient for describing the dynamics of fast gates and we find that four to six bands are required. Additionally, we achieve non-adiabatic gates by employing time-dependent lattice depths rather than using only fixed depths. The optimized control pulses not only maintain high efficacy in the presence of laser intensity and phase noise but also result in negligible inter-well couplings.
\end{abstract}

\maketitle

\section{Introduction}\label{into}

After decades of research, many physical systems have been proven capable of serving as platform for quantum information processing, including superconducting circuits \cite{Clarke:2008ul, Devoret2013, PhysRevLett.122.110501}, trapped ions \cite{PhysRevLett.74.4091, RevModPhys.75.281, HAFFNER2008155, doi:10.1126/science.1231298}, ultracold atoms \cite{RevModPhys.80.885, Saffman_2016, doi:10.1126/science.aal3837}, photons \cite{Knill2001,Prevedel2007, Flamini_2019, RevModPhys.79.135}, defects in solids \cite{doi:10.1126/science.1139831,doi:10.1073/pnas.1003052107}, and quantum dots \cite{doi:10.1103/PhysRevLett.83.4204, doi:10.1126/science.1116955, RevModPhys.79.1217}. Among these promising systems, neutral atoms have emerged over the last decade as a leading platform for quantum simulation and computing \cite{Bluvstein2024, Evered2023, PhysRevResearch.6.033104, Shaw2024}. These atoms are typically trapped in optical tweezers or lattices, allowing for the arrangement of hundreds of atoms in arbitrary geometries. 
 Quantum information is generally encoded in the internal states of the atoms, with two-qubit gates utilizing atomic interactions \cite{Jandura2022timeoptimaltwothree, PhysRevResearch.5.033052, PhysRevA.105.042430, PhysRevLett.123.170503,Goerz_2011}. These interactions can be short-ranged, such as van der Waals interactions \cite{PhysRevLett.93.063001}, or have considerably longer ranges, such as dipole-dipole interactions \cite{PhysRevLett.99.073002}. One successful approach is
excitation to Rydberg states, which facilitates strong and controllable interactions, enabling the study of many-body dynamics and the execution of quantum logic operations \cite{PhysRevA.94.032306, PhysRevA.89.032334}.

An alternative approach to Rydberg atoms is to work with ground-state atoms trapped in optical lattices or superlattices
\cite{PhysRevLett.82.1975, PhysRevLett.82.1060, PhysRevA.77.052333, doi:10.1126/science.1150841,doi:10.1126/science.aaz6801,Anderlini2007,PhysRevA.70.012306, PhysRevA.71.050701, PhysRevA.61.022304, zhu2025splittingconnectingsingletsatomic,PhysRevLett.107.165301}. A superlattice is formed by superimposing at least two optical lattices with different wavelengths and has multiple double well structures \cite{chalopin2024opticalsuperlatticeengineeringhubbard, Impertro_2024}. These systems naturally realize the Hubbard model \cite{RevModPhys.80.885, TARRUELL2018365}.  Neutral atoms in an optical lattice can be arranged in arbitrary configurations using efficient atom transport mechanisms \cite{PhysRevX.11.011035, PhysRevResearch.6.033282, cicali2024neutralatomtransporttransfer}. The exchange of atoms and their entanglement is generated through controlled collisions between the atoms, enabling the implementation of essential quantum gates such as SWAP and $\sqrt{\mathrm{SWAP}}$ \cite{Anderlini_2007, PhysRevResearch.3.013113}. Combined with single-qubit rotations, these gates constitute a universal gate set for quantum simulation and computation \cite{PhysRevLett.98.070501, Nielsen_Chuang_2010}.
The collision interactions are tuned by adjusting the barrier height of the double well and its scattering length. To ensure that the system adheres to Hubbard-model dynamics, the exchange interactions must be controlled to prevent atom excitations~\cite{PhysRevA.107.033323,PhysRevResearch.6.L042024}. Most directly,
this is achieved by adiabatically changing the barrier heights \cite{PhysRevLett.98.070501}. However, this approach results in slower gates which can reduce the coherence time and affect longer circuits in quantum simulation and computation.

Faster and more efficient gates can be achieved using optimal control methods, which have become an integral part of quantum
computing, quantum simulation, and quantum information
processing \cite{Koch2022, Glaser2015, PhysRevB.74.161307, Khani_2009, doi:10.1126/science.aax9743}. The efficiency of quantum operations is increased by shaping their driving fields using analytical
and numerical techniques. Analytical methods are usually only applicable to smaller systems requiring a detailed understanding of their dynamics \cite{PhysRevA.63.032308, PhysRevA.65.032301, PhysRevA.66.012305, PhysRevLett.88.237902, PhysRevA.70.032319, PhysRevLett.96.060503, PhysRevA.75.012322, PhysRevA.77.032332, PhysRevA.84.062301, PhysRevA.86.054302, KHANEJA2020108639}, while the results of 
\cite{Boscain-Rudolf,Yuan-Zeier-Khaneja, Khaneja-Heitmann-Glaser, Nimbalkar-Zeier-Glaser,VanDamme-Zeier-Glaser-Sugny}
are relevant for the analytical aspects of our work.
In contrast, numerical
methods can often be implemented with partial (or no) knowledge of the system dynamics
and they can depend on the system and rely on open-loop or closed-loop approaches 
(i.e.\ without or with feedback)
\cite{KHANEJA2005296, PhysRevA.92.062343, doi:10.1073/pnas.1716869115, Nigmatullin_2009, DEFOUQUIERES2011412, PhysRevA.84.022326, PhysRevLett.106.190501, Muller_2022, PhysRevA.84.022305, ROSSIGNOLO2023108782, 10.1116/5.0006785, PhysRevA.93.013851, vanFrank2014, PhysRevA.84.012312, PhysRevLett.103.110501, PhysRevA.96.042306}.

In this article, we use open-loop optimal control techniques to develop fast SWAP and $\sqrt{\mathrm{SWAP}}$ gates for fermionic $^6$Li atoms trapped in a superlattice. We describe our system with a Fermi-Hubbard model using realistic experimental parameters \cite{chalopin2024opticalsuperlatticeengineeringhubbard, Impertro_2024,doi:10.1126/science.aaz6801}. 
To this end, we optimize the lattice depths and s-wave scattering parameter to achieve high-fidelity state transfer from an initial state $\Psi_0=\ket{\ua\da}$ to target states $\Psi_{\mathrm{SWAP}}=\ket{\da\ua}$ and $\Psi_{\sqrt{\mathrm{SWAP}}}:=[(1{+}i)\ket{\ua\da} - (1{-}i)\ket{\da\ua}]/2$ and eventually find high-fidelity gates. Our study reveals that the two-band Fermi-Hubbard model is insufficient in explaining the dynamics of these fast gates (see Result~\ref{result-three}).

The key idea of our fast gates is to extend the Fermi-Hubbard model to include the higher energy bands
(see Sec.~\ref{hamiltonian_des}). 
We numerically demonstrate, using realistic experimental parameters that the control duration is as short as $0.08$~ms for transferring the state $\Psi_0$ to $\Psi_{\mathrm{SWAP}}$ with fidelity $0.999$, and $0.12$~ms for transferring $\Psi_0$ to $\Psi_{\sqrt{\mathrm{SWAP}}}$ with fidelity $0.995$, which is five times shorter than typical experimental state transfer times \cite{Impertro_2024, chalopin2024opticalsuperlatticeengineeringhubbard} (see Result~\ref{result-four}).
These fidelities can be further improved and we show that they are mainly limited by the available laser power. We also detail that these optimized controls are robust against intensity and phase noises and result in minimal inter-well tunneling (see Result~\ref{result-five}), which enables improved coherence times of 460 Rabi oscillations compared to 33 Rabi oscillations in a recent experimental study \cite{chalopin2024opticalsuperlatticeengineeringhubbard}. Moreover, we analyze in Sec.~\ref{three_four_atom}
how our gates
perform when they are applied to error states with 
three and four atoms in a double well, which provides a more complete understanding
of the system dynamics under various experimentally possible scenarios. Lastly, in Sec.~\ref{sec:full:gate}, we show that for fast gates, the initial states $\ket{\uparrow\uparrow}$ and $\ket{\downarrow\downarrow}$ can excite to higher levels, which can be minimized with a full gate optimization. The full gate optimization yields a SWAP gate with a fidelity of $0.997$ in $0.10$~ms and a $\sqrt{\mathrm{SWAP}}$ gate with a fidelity of $0.993$ for a gate duration of $0.16$~ms (see Result~\ref{full_gate_res}).

Our results initially arise from analytical and numerical optimizations using a two-band Fermi-Hubbard model
(see Results~\ref{result-one} and \ref{result-two}), where the gradients for the numerical optimization are computed based on
a particularly effective spline-fit approach (see Sec.~\ref{numeric}). For higher-band Fermi-Hubbard models, the spline-fit
approach is not applicable in the presence of multiple hopping and interaction terms, 
and we instead develop an approach using an effective approximate analytical gradient in Appendix.~\ref{approximate-analytic-gradient}.
The performance of the different gradient approaches is compared in Appendix~\ref{sec:gradient:methods}.
Thus our work combines faster, robust two-qubit gates for ultracold atoms
under realistic experimental conditions with methodological advances.

The structure of the paper is as follows: In Sec.~\ref{Two-band-Fermi-Hubbard}, we define the Hamiltonian for the superlattice potential, introduce the corresponding two-band Fermi-Hubbard model and explain the objective of the quantum control tasks. We then optimize the SWAP and $\sqrt{\mathrm{SWAP}}$ gates using both analytical and numerical techniques in Sec.~\ref{optimization with two band}. In Sec.~\ref{Excitation to higher bands}, we explore the impact of higher energy bands on the optimized fast gates and demonstrate the limitations of the two-band Fermi-Hubbard model. Next, in Sec.~\ref{optimization with higher band}, we present the optimization of the SWAP and $\sqrt{\mathrm{SWAP}}$ gates using the higher-band Fermi-Hubbard model. In Sec.~\ref{three_four_atom}, we investigate the dynamics of more than two atoms in a double well under the optimal control pulses, and analyze the robustness of the optimized control pulses in Sec.~\ref{subsec:robustness}. We perform the full gate optimization to minimize the gate error for different basis states in Sec.~\ref{sec:full:gate}.
In Sec.~\ref{sec:high:interaction},
we discuss the effect of high interaction strength on $\sqrt{\mathrm{SWAP}}$ transfer.
Finally, we summarize our conclusions in Sec.~\ref{Conclusion}. The raw data files from the
simulations performed for this work are provided in Ref.~\cite{JUELICH-DATA/GG80VM_2025}.

\section{Model and objective\label{Two-band-Fermi-Hubbard}}
\subsection{Model}\label{model}
\begin{figure}
    \includegraphics{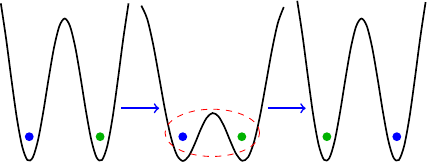}
    \caption{Collision gate in a double well. A double well is initialized with two $^6$Li atoms, where the green one on the right
    is in the state $\ket{\ua}$ and the blue one on the left is in the state $\ket{\da}$ which yields the state $\ket{\ua\da}$.
    The barrier inside the double well is lowered in a controlled way to allow the two atoms to interact. This results in a SWAP or a $\sqrt{\mathrm{SWAP}}$ gate between the two atoms. The SWAP gate exchanges the spins and prepares the state $\ket{\da\ua}$, while the $\sqrt{\mathrm{SWAP}}$ gate prepares the state $[(1{+}i)\ket{\ua\da} - (1{-}i)\ket{\da\ua}]/2$.
    \label{fig:schematic_2}}
\end{figure}

We consider a double well potential, barrier heights of which can be controlled dynamically \cite{chalopin2024opticalsuperlatticeengineeringhubbard, Impertro_2024}, as
\begin{align}
V(\mathbf{r},t)=& V_s (t)\cos^2 (k_s x {+}\varphi(t)) - V_\ell(t) \cos^2 (k_{\ell} x)  \nonumber \\
&-V_y \cos^2 (k_y y )-V_z \cos^2 (k_z z )
\label{potential_2}
\end{align}
where $\mathbf{r}=(x,y,z)$.
In the $x$ direction, a superlattice or a double well potential [see Fig.~\ref{fig:schematic_2}] is created by standing waves from two tilted
lasers with a short wavelength of $\lambda_s{=}532$~nm and two lasers with a long
wavelength of $\lambda_\ell{=}1064$~nm. The wave vectors of the lattice are then given by 
$k_{b}{=}{2 \pi}/{\lambda_{b}}\sin({26.7^\circ}/{2})$ with $b \in \{s,\ell\}$
where $26.7^\circ$ is the angle between the tilted lasers constructing the lattice, chosen for the particular geometry of the experimental system \cite{chalopin2024opticalsuperlatticeengineeringhubbard}. 
Here, $V_s(t)$ and $V_\ell(t)$ respectively denote the 
tunable lattice depths for the short and long lattice, where the short lattice is blue detuned (repulsive) and the long lattice is
red detuned (attractive). The tunable relative phase
between the short and long lattice is given by $\varphi(t)$. We consider
optical lattices with constant lattice depths $V_y = 45 E_{ry}$ and $V_z =45 E_{rz}$ in the $y$ and $z$ direction, expressed in units of their respective recoil energies $E_{r{y}}{=}\hbar^2k_{y}^2/(2\mass)$
and $E_{r{z}}{=}\hbar^2k_{z}^2/(2\mass)$. 
Following the experimental work of \cite{chalopin2024opticalsuperlatticeengineeringhubbard},
we initialize the double well with two fermionic $^6$Li atoms, where one atom is in the spin-up state 
and the other one in the spin-down state as shown in Fig.~\ref{fig:schematic_2}.
We model the system of two atoms based on the Hamiltonian
\begin{align}
 H_1(\mathbf{r},t)&=-\frac{\hbar^2}{2\mass} \nabla^2+ V(\mathbf{r},t) \label{real_space_ham_1}, \\
H_2(\mathbf{r}_1,\mathbf{r}_2,t)&=\sum_{j=1}^2 H_1(\mathbf{r}_j,t) +U_{3 \mathrm{D}}(\mathbf{r}_1,\mathbf{r}_2),
\label{real_space_ham_2}
\end{align}
where $\mass$ is the mass of the $^6$Li atoms and $H_1$ is the single-atom Hamiltonian. Here, 
the pseudo interaction potential $U_{3 \mathrm{D}}(\mathbf{r}_1,\mathbf{r}_2)$ between the atoms is defined as
\begin{equation}
U_{3 \mathrm{D}}(\mathbf{r}_1,\mathbf{r}_2)=\frac{4 \pi \hbar^2}{\mass} a \delta(\mathbf{r}_1{-}\mathbf{r}_2),
\label{u3d}
\end{equation}
and $a$ denotes the characteristic scattering length. In an experiment, $a$ is tuned by changing the magnetic field, giving Feshbach resonances \cite{RevModPhys.82.1225}.
Since the optical potential of Eq.~\eqref{potential_2} separates into the three spatial components, one
can independently solve 
them in the non-interacting case.
Thus we can concentrate only on the $x$ direction dependence of Eq.~\eqref{real_space_ham_1} and Eq.~\eqref{real_space_ham_2} for
our calculations. For interacting particles, this argument holds approximately if one assumes that either the control pulses remain
in the adiabatic regime (without excitations in the $y$ and $z$ directions) or the optical potential in the $y$ and $z$ directions 
is strong enough
such that the interaction energy is low enough to not excite the atoms in those dimensions.
A mixture of both assumptions is valid for our work.
We write the single-atom Hamiltonian of Eq.~\eqref{real_space_ham_1} in Fourier space as
\begin{equation}\label{fourier_ham}
\tilde{H}_1(q)=\frac{\hbar^2}{2\mass} q^2 + \tilde{V}(q),
\end{equation}
where $\tilde{V}(q)$ is the Fourier expansion of the lattice potential with quasi-momentum $q$. We assume a lattice with $L$ sites and discretized values for the momenta $q=f k_s +k$ with $f \in \mathbb{Z}$,
 $k= k_s ({2n{+}1{-}L})/{2L}$, and $n \in [0,L{-}1]$ in the first Brillouin zone.

\begin{figure}
\includegraphics{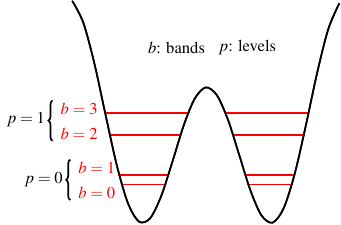}
\caption{Level scheme of the multi-band Fermi-Hubbard model for one double well of the superlattice potential described by Eq.~\eqref{potential_2}.
The bands appear in pairs, with each pair corresponding to one level $p$ of the model. 
For instance, four bands form two levels, where the zeroth ($b=0$) and 
first ($b=1$) bands form level $p=0$, and the second ($b=2$) and third ($b=3$) bands form level $p=1$.\label{fig:schematic_3}} 
\end{figure}

\begin{figure}
\includegraphics{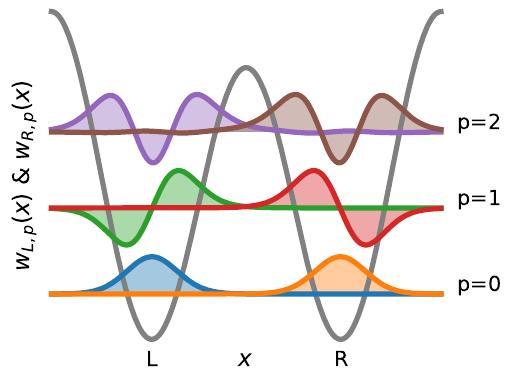}
\caption{The Wannier states $w_{L,p}$ and $w_{R,p}$ in the left and the right bases of the first three levels $p=0,\, p=1$, and $p=2$
for the corresponding optical potential in the $x$ direction. Each level $p$ is constructed by a linear combination of the two bands $b=2p$ and $b=2p+1$ as shown also in Fig.~\ref{fig:schematic_3}.}
\label{wannier_states}
\end{figure}

From Bloch's theorem, the eigenfunctions of $\tilde{H}_1(q)$ can be written in the form 
\begin{align*}
&\Phi_{b,k}(x)=e^{ikx/\hbar}\, \phi_{b,k}(x)\\
&\text{where}\; \phi_{b,k}(x)= \sum_m C_{m,b,k}\, e^{ifk_{s}x} .
\end{align*}
Here, $C_{m,b,k}$ are coefficients obtained by solving for the eigenstates of Eq.~(\ref{fourier_ham}),
the index $b$ denotes the band index, and $\phi_{b,k}(x)$ are periodic functions with periodicity $d=\lambda_\ell /2$, i.e.,
$\phi_{b,k}(x)=\phi_{b,k}(x{+}d)$.
Bloch states $\Phi_{b,k}(x)$ are localized in momentum space, but completely delocalized in real space. Equivalently, we can define a set of basis states called Wannier states, which are localized in real space \cite{PhysRev.52.191, PhysRev.115.809, PhysRevB.7.4388}.
We choose the Wannier states as our basis in which the system is specified.
This is motivated as we assume the wave function of the atoms 
to be localized on one side of the double well and close to the ground state of the optical lattice. 
Under these assumptions, the wave functions can be described with only a few Wannier states.
We can calculate Wannier functions $w_b$ based on the idea that these states are defined as the eigenstates
of the position operator $\mathrm{X}$ such that
\begin{equation*}
\mathrm{X}_b\, w_b(x{-}x_0)=x_0\, w_b(x{-}x_0).
 \end{equation*}
For this, we write the position operator 
\begin{equation*}
\mathrm{X}_{b,k',k} =\int \Phi_{b,k'}^*(x)\, x\, \Phi_{b,k}(x) dx
 \end{equation*}
in terms of the Bloch state basis $\Phi_{b,k}(x)$.
The Wannier function $w_b(x{-}x_0)$ in band $b$ at position $x-x_0$ in the $x$ direction  is defined as
\begin{equation*}
 w_b(x{-}x_0)=\frac{1}{\sqrt{L}}\sum_k e^{-i(kx_0+\chi_k)}\,\Phi_{b,k}(x)
 \end{equation*}
 with a momentum-dependent phase $\chi_k$, which is given by the condition that the Wannier states are the eigenstates of the position operator $\mathrm{X}$.
Note that even if the Wannier states build an orthonormal basis, they are not the eigenstates of the Hamiltonian. 
Nevertheless, the stronger the optical potential gets, the flatter the bands become in $k$, and with that Wannier states match closer with the eigenstate of the Hamiltonian. 
For a double well system as shown in Fig.~\ref{fig:schematic_3}, 
one can always take a linear combination of two bands ($2p$ and $2p+1$) to construct the left $w_{L}$ and right $w_{R}$ basis states of the level $p$.
For that, one only needs to apply the condition that the Wannier states $w_{L}$ and $w_{R}$ are still the eigenstates of the position operator $\mathrm{X}$. Examples of Wannier states are shown in Fig.~\ref{wannier_states}.

\begin{table}
\centering
\begin{tabular}{@{\hspace{1mm}} c @{\hspace{4mm}} c @{\hspace{4mm}} c @{\hspace{4mm}} c @{\hspace{1mm}} }
\hline\hline
\\[-3mm]
Model&Gate & Fermi-Hubbard & Experimental\\
&& parameters & parameters\\[1mm]
\hline
\\[-2mm]
Two-band&SWAP& $J(t)$  &$V_s(t)$, $V_\ell(t)$\\
\\[-1mm]
 & $\sqrt{\mathrm{SWAP}}$&$J(t)$&$V_s(t)$, $V_\ell(t)$ \\[1mm]
&&$U$ & $a$\\
 \\
 Higher-band&SWAP& $J_p(t)$ &$V_s(t)$, $V_\ell(t)$\\
 \\[-1mm]
 & $\sqrt{\mathrm{SWAP}}$&$J_p(t)$&$V_s(t)$, $V_\ell(t)$ \\[1mm]
&&$U_{mnop}^{\alpha\beta\gamma\delta}(t)$ & $a$, $V_s(t)$, $V_\ell(t)$\\[1mm]
\hline\hline
\end{tabular}
\caption{Different Fermi-Hubbard parameters and the corresponding tunable experimental parameters.
There are two Fermi-Hubbard parameters: interaction strength $U$ and hopping strength $J$.
For the two-band Fermi-Hubbard model, $J$ is time-dependently tuned using the lattice depths $V_s$ and $V_\ell$. Whereas, we assume that $U$ is time-independent and tuned by the scattering length $a$.
The conversion relations between the experimental parameters and the Fermi-Hubbard parameters are given by Eq.~\eqref{v_to_j}-\eqref{eqn_for_U}.
For the higher-band model, we have multiple $J_p$ for different $p$ levels
of the double well controlled by $V_s$ and $V_\ell$ with Eq.~\eqref{v_to_j_general}. 
Similarly, we have multiple time-dependent $U_{mnop}^{\alpha\beta\gamma\delta}(t)$ tuned by $a$, $V_s$ and $V_\ell$ with Eq.~\eqref{eqn_for_U_general}
as detailed in Sec.~\ref{hamiltonian_des}.}
\label{table:table_parameter}
\end{table}

Instead of simulating the real-space Hamiltonian of Eq.~\eqref{real_space_ham_2}, we can also describe fermionic atoms trapped in the optical lattices with the Fermi-Hubbard model \cite{doi:10.1146/annurev-conmatphys-070909-104059, PhysRevLett.89.220407} as
\begin{equation}
\hat{H}=-J\sum_{i\neq j,\sigma}(c_{i\sigma}^{\dagger}c_{j\sigma}{+}h.c.)
+U\sum_{i}\mathfrak{n}_{i\uparrow}\mathfrak{n}_{i\downarrow},
\label{fermi-hubbard-model}
\end{equation}
where the operator $c_{i\sigma}^{\dagger}$ (or $c_{i\sigma})$ creates (or annihilates) an atom in the spin state $\sigma \in \{\ua,\da\}$
at the lattice site $i$. The operators $c_{i\sigma}^{\dagger}$ and $c_{j\sigma'}$ have anti-commutation relation as 
$\{c_{i\sigma}^{\dagger}, c_{j\sigma'}\}=\delta_{ij}\delta_{\sigma \sigma'}$, and
$\mathfrak{n}_{i\sigma} = c_{i\sigma}^{\dagger}c_{i\sigma}$ is the number operator. 
The first term represents the hopping between site $i$ and $j$ of the lattice with amplitude $J$. The second term describes 
the interaction of strength $U$ between two atoms of opposite spins, 
sitting on the same site $i$ of the lattice.
The model of Eq.~\eqref{fermi-hubbard-model} is sufficient to describe the atoms in the optical lattice with periodic single wells. However, for a superlattice with periodic double wells and relevant depth, the gap between the two lowest energy bands becomes very small inside each double well as shown in Figs.~\ref{fig:schematic_2} and \ref{fig:schematic_3}. Therefore, we extend the one-band Fermi-Hubbard model to a two-band model for describing
the dynamics of the double well.
We assume that all the higher bands ($b > 1$) as shown in Fig.~\ref{fig:schematic_3} are 
still separated by a large energy gap and the two-band model 
sufficiently describes the system.

The two-band Fermi-Hubbard model for two atoms in a single double well \cite{Impertro_2024, chalopin2024opticalsuperlatticeengineeringhubbard} is given by
\begin{equation}
\hat{H}=-J\sum_{\sigma}(c_{L\sigma}^{\dagger}c_{R\sigma}{+}h.c.)
+U\hspace{-2mm}\sum_{\alpha=L,R}\mathfrak{n}_{\alpha\uparrow}\mathfrak{n}_{\alpha\downarrow}\; ,
\label{fermi-hubbard-model-two}
\end{equation}
where $c_{L\sigma}^{\dagger}$ (or $c_{L\sigma}$) and $c_{R\sigma}^{\dagger}$ (or $c_{R\sigma}$) 
create (or annihilate) a fermion in spin state $\sigma \in \{\uparrow,\downarrow\}$ on the left (L) or the right (R) side of
the double well respectively. Equation~\eqref{fermi-hubbard-model-two} resembles Eq.~\eqref{fermi-hubbard-model} with
the site index $i$ replaced by the left (L) and right (R) side of the double well.
The first term represents the hopping between the left and the right side of the double well 
with amplitude $J$. The second term describes 
the interaction of strength $U$ between two atoms of opposite spins, 
sitting on the same side of the double well. Throughout the paper, we work with a symmetric double well,
i.e.\ $\varphi(t){=}0$ [see Eq.~\eqref{potential_2}]. Thus $U$ is identical for the left and right wells.
The gates are based on the hopping and the onsite interaction of the atoms in the double well, which are given by 
\begin{align}
J&=-\int w_R(x)H_1(x)w_L(x) dx  \label{v_to_j}, \\
U&= \iint U_{3 \mathrm{D}}(\mathbf{r}_1,\mathbf{r}_2)\, w_{\alpha}^2(\mathbf{r}_1)\, w_{\alpha}^2(\mathbf{r}_2)\, d\mathbf{r}_1 \, d\mathbf{r}_2 \label{eqn_for_U},
\end{align}
where $U$ is assumed to be independent of $\alpha\in \{L, R\}$.
The pseudo interaction potential $U_{3 \mathrm{D}}(\mathbf{r}_1,\mathbf{r}_2)$ is defined in Eq.~\eqref{u3d} and $H_1(x)$ is the single-atom Hamiltonian from Eq.~\eqref{real_space_ham_1} in the $x$ direction. The function $w_{\alpha}(\mathbf{r})=w_{\alpha}(x)w_0(y) w_0(z)$ is the three-dimensional Wannier state, where $w_{L}(x)$ and $w_{R}(x)$ represent the states in the $x$ direction on the left (L) and the right (R) side of the double well, respectively. The Wannier states of the lowest band in the $y$ and the $z$ 
direction are $w_0(y)$ and $w_0(z)$, respectively. 
As explained in Table~\ref{table:table_parameter}, the hopping strength $J$ is calculated from the 
time-dependent lattice depths $V_s(t)$ and $V_\ell(t)$. The onsite interaction $U$ depends on the constant scattering length $a$, along with the lattice depths $V_s(t)$ and $V_\ell(t)$. In the two-band description, $U$ is almost independent of the change in the long lattice depth $V_\ell$ and it increases with increasing short lattice depth $V_s$. However, $U$ can be tuned over a much larger range by changing $a$, $V_y$, and $V_z$ compared to changing $V_s$ \cite{Frederik_thesis}. As a simplification, for the two-band model, we assume that the change in the onsite interaction depends only on the change in the s-wave scattering with tunable constant scattering length $a$. 
We can calculate $J(t)$ and $U$ from the experimental parameters $V_s, V_\ell$ and $a$ using Eqs.~\eqref{v_to_j} and \eqref{eqn_for_U} respectively. However, the reverse calculation of the experimental parameters $V_s(t)$ and  $V_\ell(t)$ from a given $J(t)$ is more problematic as explained in Sec.~\ref{numeric}.
\subsection{Objective}\label{objective}
The two-qubit gates for the fermionic system in the superlattice can be represented using the spin non-conserving basis states $\ket{\ua\ua}$, $\ket{\ua\da}$, $\ket{\da\ua}$, and $\ket{\da\da}$, 
where each state represents atoms on the left (L) and the right (R) side of the double well \cite{PhysRevResearch.3.013113, PhysRevLett.98.070501}.
So, the state
$\ket{\ua\da}$ describes that the first atom in the left (L) well
is in the spin-up state 
and the second atom in the right (R) well observes a spin-down state. In this basis, the total spin of the system is not 
conserved. The SWAP and $\sqrt{\mathrm{SWAP}}$ gates are represented in this basis as
\begin{equation*}
\mathrm{SWAP} =
\left[
\begin{smallmatrix}
1 & 0 & 0&0  \\
0 & 0 &1& 0  \\
0&1&0 & 0\\
0&0&0 & 1\\
\end{smallmatrix}\right],\;
 \sqrt{\mathrm{SWAP}}=
 \left[
 \begin{smallmatrix}
1 & 0 & 0&0  \\
0 & (1+i)/{2} &(-1+i)/{2}& 0  \\
0&(-1+i)/{2}&(1+i)/{2} & 0\\
0&0&0 & 1\\
\end{smallmatrix}\right].
\end{equation*}
In the two-band Fermi-Hubbard model, two fermionic atoms with the spin up or down can only attain the state $\ket{\ua\ua}$ or $\ket{\da\da}$ respectively, and hence are 
fixed by Pauli's exclusion principle. 
Therefore, applying a SWAP or  $\sqrt{\mathrm{SWAP}}$ gate will not change their states. Therefore, we focus on the basis $\ket{\ua\da}$ and $\ket{\da\ua}$. 
From the symmetry of the Hamiltonian, SWAP or  $\sqrt{\mathrm{SWAP}}$ gate will perform the same operation on $\ket{\ua\da}$ and $\ket{\da\ua}$,
which means any pulses transferring the state $\ket{\ua\da}$ to $\ket{\da\ua}$ will also transfer $\ket{\da\ua}$ to $\ket{\ua\da}$. 
Hence, it is sufficient to optimize the transfer of the 
state 
\begin{equation*}
\Psi_{0}:=\ket{\ua\da} \;\text{ to }\; \Psi_{\mathrm{SWAP}}:=\ket{\da\ua}
\end{equation*}
using a time-dependent control $J(t)$ at $U=0$ and later verify that this also transfers $\ket{\da\ua}$ to $\ket{\ua\da}$. 
Similarly, it is enough for the optimization of the $\sqrt{\mathrm{SWAP}}$ gate to optimize the transfer of the state 
\begin{equation*}
\Psi_{0} \;\text{ to }\;
\Psi_{\sqrt{\mathrm{SWAP}}}:=[(1{+}i)\ket{\ua\da} - (1{-}i)\ket{\da\ua}]/2
\end{equation*}
using $J(t)$ and the time-independent control $U$.
Therefore, we only need to solve a state-to-state transfer problem
for obtaining the SWAP and $\sqrt{\mathrm{SWAP}}$ gates in the two-band Fermi-Hubbard model. 

With one atom in the spin-up state and a second one in the spin-down state, we can have four possible states
$\ket{D0}$, $\ket{\ua\da}$, $\ket{\da\ua}$, and $\ket{0D}$, where
$D$ denotes double occupancy on the left or the right side of the double well. The states $\ket{D0}$ and $\ket{0D}$ are out of our two-qubit computational basis. Hence, our goal is to minimize the probability that the states $\ket{D0}$ and $\ket{0D}$ are observed at the end of the SWAP or $\sqrt{\mathrm{SWAP}}$ gates.
Additionally, the double well can accommodate up to four fermionic atoms with two atoms in the spin-up state and two atoms in the spin-down state. One-atom states serve as single-qubit states. 
However,
for two-qubit gates, states with three and four atoms are error states. Hence, we do not consider these error states in the optimization and only check the performance of our optimized pulses for these error states in Sec.~\ref{three_four_atom}.

\section{Optimization with a two-band Fermi-Hubbard model\label{optimization with two band}}
As explained in Sec.~\ref{objective}, we aim to optimize the transfer from $\Psi_0$ to $\Psi_{\mathrm{SWAP}}$ or $\Psi_{\sqrt{\mathrm{SWAP}}}$ by minimizing the population of the state $\ket{D0}$ and $\ket{0D}$. 
First, we optimize the hopping parameter $J(t)$ for the SWAP gate using 
analytical methods. Later, we compare our results with the numerical optimization in Sec.~\ref{numeric}.

\subsection{Analytical optimization\label{analytical}}
In this section, we study the problem of finding the time-optimal pulse sequence to perform the SWAP gate. In particular, we find the minimum time for transferring the state $\ket{\ua\da}$ to $\ket{\da\ua}$
in the two-band Fermi-Hubbard model
without any interaction, i.e., \ for $U=0$. The interaction strength $U$ can, in principle, be tuned to zero either by changing the internal atomic state to a non-interacting one or by tuning the magnetic field to set $a=0$~\cite{ediss19440,ediss20229}. The time optimal pulse also transfers the state $\ket{\da\ua}$ to $\ket{\ua\da}$. 
Similar problems have been extensively studied in other systems \cite{Boscain-Rudolf,Yuan-Zeier-Khaneja, Khaneja-Heitmann-Glaser, Nimbalkar-Zeier-Glaser,VanDamme-Zeier-Glaser-Sugny}.   
The system Hamiltonian is given by
\begin{align}
H'(t)&=\begin{bmatrix}
U&-J(t)&-J(t)&0  \\
-J(t) & 0 & 0&-J(t)  \\
-J(t)& 0&0&-J(t) \\
0&-J(t)&-J(t)&U  \\
\end{bmatrix}
\label{full_two_band_matrix}
\end{align}
in the basis corresponding to the basis states $\ket{D0}$, $\ket{\ua\da}$, $\ket{\da\ua}$, $\ket{0D}$.
We switch to a new basis $(\ket{D0} {+}\ket{0D})/\sqrt{2}$, $\ket{\ua\da}$, $\ket{\da\ua}$, $(\ket{D0} {-}\ket{0D})/\sqrt{2}$
which allows us to neglect the state 
$(\ket{D0} {-}\ket{0D})/\sqrt{2}$ (in our analytic optimization)
as it does not couple with the other states.
Therefore, the time-evolution with $\tilde{J}(t){=}{-}\sqrt{2}J(t)$ and $U{=}0$ is
\begin{equation}\label{eqn_for_motion_mat}
i\begin{pmatrix}
\dot{x}_1\\
\dot{x}_2\\
\dot{x}_3
\end{pmatrix}
=
\begin{bmatrix}
0 & 0 & \tilde{J}(t)  \\
0 & 0 & \tilde{J}(t)  \\
\tilde{J}(t)& \tilde{J}(t) & 0 \\
\end{bmatrix}
\begin{pmatrix}
x_1\\
x_2\\
x_3
\end{pmatrix}
\end{equation}
where $x_1, x_2$, and $x_3$ correspond to complex coefficients of the states $\ket{\ua\da}$,  $\ket{\da\ua}$, and $(\ket{D0} {+}\ket{0D})/\sqrt{2}$ respectively. Thus, optimizing the SWAP gate reduces to finding $\tilde{J}(t)$ for evolving the system
from $(1,0,0)^T$ to $(0,1,0)^T$. 
In order to obtain the time-optimal solution, we follow the standard approach
\cite{Boscain-Rudolf,Yuan-Zeier-Khaneja, Khaneja-Heitmann-Glaser, Nimbalkar-Zeier-Glaser,VanDamme-Zeier-Glaser-Sugny}
which is based on solving suitable Euler-Lagrange equations. 
The calculation is detailed in Appendix~\ref{app:analytic} and shows that time-optimal $J(t)$ is a constant pulse.
Other more direct
approaches might be applicable in this particular case, but the calculations in Appendix~\ref{app:analytic} also prepare the ground
for future work to explore analytical solutions beyond the considered two-band Fermi-Hubbard model.

\begin{res}\label{result-one}
The time optimal SWAP gate in the two-band Fermi-Hubbard model without any interaction (i.e.\ $U=0$)
is given by a constant pulse $J(t)=J_\text{max}$, where $J_\text{max}$ is the maximal experimentally possible hopping strength $J$.  The minimum gate duration to transfer the state from $\ket{\ua\da}$ to  {$-\ket{\da\ua}$} or vice-versa is given by $T{=}{\pi}/{(2J_\text{max})}$. For $J_\text{max}=34.03$~kHz, we get $T{=}{\pi}/{(2J_\text{max})}=0.046$~ms. The explicit
form of the state evolution follows from Eq.~\eqref{analytic_soln}.
\end{res}

\begin{rem}\label{phase_rem}
The state $\ket{\ua\da}$ can only be transformed into ${-}\ket{\da\ua}$ using a real $J(t)$. The $-1$ phase of $\ket{\da\ua}$ appears since
the Hamiltonian is written using the spin-ordered convention where the creation operators are applied on the vacuum $\ket{00}$ in the order $c^\dagger_{R\da}c^\dagger_{L\da}c^\dagger_{R\ua}c^\dagger_{L\ua}$ and thus 
\begin{equation*}
\mathrm{{SWAP}'} =
\left[
\begin{smallmatrix}
1 & 0 & 0&0  \\
0 & 0 &-1& 0  \\
0&-1&0 & 0\\
0&0&0 & 1\\
\end{smallmatrix}\right].\;
\end{equation*}
We can also write the Hamiltonian using the site-ordered convention where the creation operators are applied on the vacuum $\ket{0}$ in the order $c^\dagger_{R\da}c^\dagger_{R\ua}c^\dagger_{L\da}c^\dagger_{L\ua}$. The site-ordered convention gives the usual SWAP operation. 
One can apply the matrix $\mathrm{diag}(1,-1,-1,1)$ on $\mathrm{{SWAP}'}$ to get $\mathrm{{SWAP}}$, which requires a phase change conditional on the spin parity. Optimizations performed in this work uses the spin-ordered convention, but the results are easily adaptable to site-ordered convention. 
Therefore, we neglect the phase of the target states for our numerical optimizations.
\end{rem}
In the next section, we use a gradient-based numerical optimization to find the optimal $J(t)$
for the SWAP gate and compare it with the analytical result.

\begin{figure}
\includegraphics{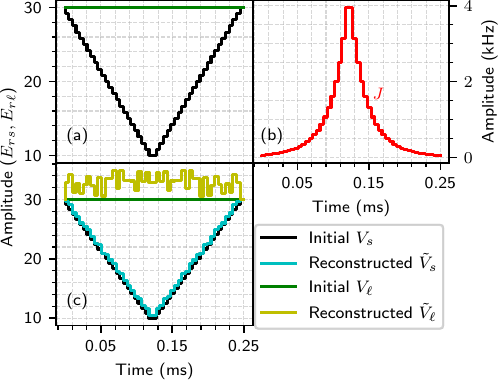}
\caption{Relation between (a) lattice depths $V_s$ and $V_\ell$
in their respective recoil
energies $E_{r{s(\ell)}}{=}\hbar^2k_{s(\ell)}^2/2m$
and (b) the corresponding hopping strength $J$ via Eq.~\eqref{v_to_j}.
(c)~The non-recommended noisy reconstruction of $V_s$ and $V_\ell$
from the hopping strength $J$ in (b) significantly differs from 
their initial values in (a). This is avoided by directly
optimizing $V_s$ and $V_\ell$ as detailed in Sec.~\ref{numeric}.
\label{J_to_V}}
\end{figure}

\begin{algorithm}[t]
\caption{Optimization of the SWAP and $\sqrt{\mathrm{SWAP}}$ gates using the two-band model}\label{optimization_algorithm_two_band}
\begin{algorithmic}
\State \textbf{SWAP}:
\State \textbf{Input}: $a=0$
\State \textbf{Optimization parameters}: $V_s(t)$ and $V_\ell (t)$
\State \textbf{Step 1}: Define the SWAP-gate cost function $C_{\mathrm{SWAP}}=1- \abst{{\langle \Psi_{\mathrm{SWAP}} \vert \Psi(T) \rangle}}^2$
\State \textbf{Step 2}: Minimize $C_{\mathrm{SWAP}}$ to find optimal $V_s$ and $V_\ell$ using gradient-based optimization with the spline-fit method
\State \textbf{Output}: Optimized $V_s(t)$ and $V_\ell (t)$ for the SWAP gate
\vspace{0.3cm}
\State $\mathbf{\sqrt{\mathrm{\mathbf{SWAP}}}}$:
\State \textbf{Input}: $V_s(t)$ and $V_\ell (t)$ from the SWAP gate
\State \textbf{Optimization parameters}: $a$ 
\State \textbf{Step 1}: Define the $\sqrt{\mathrm{SWAP}}$-gate cost function $C_{\sqrt{\mathrm{SWAP}}}=1- \abst{{\langle \Psi_{\sqrt{\mathrm{SWAP}}} \vert \Psi(T) \rangle}}^2$
\State \textbf{Step 2}: Perform a one-dimensional search to minimize $C_{\sqrt{\mathrm{SWAP}}}$ and optimize $a$
\State \textbf{Output}: Optimized $a$ for the $\sqrt{\mathrm{SWAP}}$ gate
\end{algorithmic}
\end{algorithm}

\subsection{Numerical optimization\label{numeric}}
As explained in Sec.~\ref{model}, the hopping parameter $J$ is tuned by changing 
the lattice depths $V_s$ and $V_\ell$, while the interaction 
strength $U$ is controlled by the constant scattering length~$a$.
The conversion relations between the experimental parameters and the Fermi-Hubbard
parameters are given by Eqs.~\eqref{v_to_j}-\eqref{eqn_for_U}.
To perform experiments with optimal control pulses, 
we need to convert the optimized $J$ and $U$ to the experimental 
parameters $V_s$, $V_\ell$, and $a$. However, reconstructing $V_s$ and 
$V_\ell$ from $J$ is non-trivial and introduces noise. To test the reconstruction performance, we 
start with an initial set of $V_s$ and $V_\ell$ [shown in Fig.~\ref{J_to_V}(a)], 
expressed in units of their respective recoil energies $E_{r{s}}{=}\hbar^2k_{s}^2/(2\mass)$
and $E_{r{\ell}}{=}\hbar^2k_{\ell}^2/(2\mass)$.
These lattice depths generate an exponentially changing $J$, as shown in Fig.~\ref{J_to_V}(b). 
We then attempt to reconstruct the initial lattice depths from $J$ using 
a pre-stored data table of $V_s, V_\ell$, and $J$. As shown in Fig.~\ref{J_to_V}(c), 
the reconstructed long lattice depth $\tilde{V}_\ell$ does not match the initial $V_\ell$, 
and the back conversion also introduces noise in the reconstructed short lattice depth $\tilde{V}_s$.
\begin{rem}
Optimizing $J$ and converting it back to lattice depths $V_s$ and $V_\ell$
is not effective for obtaining smooth and realistic pulses. 
Therefore, for the remainder of this work, 
we directly optimize the experimental parameters $V_s$, $V_\ell$, and $a$ to achieve the optimized SWAP and $\sqrt{\mathrm{SWAP}}$ gates.
\end{rem}
We can apply a gradient-based optimization technique known as
GRAPE \cite{KHANEJA2005296} which can also utilize
Newton or quasi-Newton (BFGS) methods \cite{doi:10.1137/0916069, Byrd1994, PhysRevA.102.042612, PhysRevA.84.022307, DEFOUQUIERES2011412} and can include different transfer functions \cite{PhysRevApplied.19.064067,rasulov2025instrumentaldistortionsquantumoptimal} to optimize $V_s$ and $V_\ell$ for the SWAP gate with $a=0$.
For the $\sqrt{\mathrm{SWAP}}$ gate, we use the optimized $V_s$ and $V_\ell$ from the SWAP gate and
conduct a one-dimensional search with the target state
$\Psi_{\sqrt{\mathrm{SWAP}}}$ to identify the 
optimal scattering length $a$ for the $\sqrt{\mathrm{SWAP}}$ gate.
Algorithm~\ref{optimization_algorithm_two_band} briefly explains the optimization steps for
the SWAP and $\sqrt{\mathrm{SWAP}}$ gates for a two-band Fermi-Hubbard model.

We need to calculate gradients of the cost function with respect to $V_s$ and $V_\ell$ for the gradient-based optimization of the SWAP gate.
For this, we can express the Hamiltonian of Eq.~\eqref{eqn_for_motion_mat} as a control Hamiltonian given by 
\begin{equation}\label{control_ham}
H(t) =  {J(t)\, H_J},
\end{equation}
where $H_J$ is constant in time  and $U=0$ for the SWAP gate. 
Now, our goal is to transfer
a quantum system from the given initial pure state $\Psi_{\text{ini}}=\Psi_{0}$
to the target pure state $\Psi_{\text{tar}}=\Psi_{\mathrm{SWAP}}$ in time $T$ by varying the control
pulse $J(t)$ while minimizing the cost function
\begin{equation}\label{cost_function}
C= 1- \abst{\langle {\Psi_{\text{tar}}} \vert \Psi(T) \rangle}^2.
\end{equation}
We divide the total control duration $T$ into $N_T$ equal steps of duration $\Delta t= T/N_T$,
which results in a piecewise constant $J(t)$.
The time evolution of the quantum system during the $j$th time step is given by 
\begin{equation}\label{time_evol_jth}
\mathcal{U}_j=\exp[-i\Delta t{J({j})H_J}].
\end{equation}
The cost function \eqref{cost_function} can be written as
\begin{equation}\label{eqn:A.2}
C= 1- \abst{\langle \Psi_{\text{tar}} \vert\, \mathcal{U}_{N_T}\!\cdots \mathcal{U}_{1} \vert \Psi_\text{ini}\rangle}^2
\end{equation}
To minimize $C$, at every iteration of the algorithm, we update the controls by
\begin{equation*}\label{eqn:A.8}
J({j})\rightarrow J({j}) - \mathfrak{e}\frac{\delta C}{\delta J({j})},
\end{equation*}
where $\mathfrak{e}$ is a small unitless step matrix. 

Now, in our case, we perform a change
of controls from the experimental parameters $V_k {\in} \{V_s, V_\ell \}$ to the Fermi-Hubbard parameter $J$, 
where $V_k$ and $J$ are piecewise constant with $N_T$ time steps. 
Hence, we follow the derivation in \cite{PhysRevA.84.022307}, where
the product rule is applied to the gradient calculation and one obtains
\begin{equation*}
\frac{\delta C}{\delta V_{k}(j)}=\sum_{s=1}^{N_T}{\frac{\delta J(s)}{\delta V_k{(j)}}\frac{\delta C}{\delta J(s)}}.
\end{equation*}
The derivative ${\delta C}/{\delta J(s)}$
is here calculated with the help of the Fréchet-derivative method \cite{doi:10.1137/080716426} using the Python
package SciPy~\cite{2020SciPy-NMeth}.
We use a spline-fit method for calculating the Fermi-Hubbard parameter $J$ 
and its gradient with respect to the lattice depths at each time step and every 
optimization iteration, i.e., $\delta J(s)/\delta V_k{(j)}$. For this, we create a grid of $100\times 100$ pairs $(V_s, V_\ell)$, where $V_s$ ranges 
from 2$E_{rs}$ to 30$E_{rs}$ and $V_\ell$ ranges from 30$E_{r\ell}$ to 50$E_{r\ell}$.
We calculate and store the value of $J$ for each point on the grid.
Using the pre-stored data set of triples $(V_s, V_\ell, J)$, 
we fit a bivariate spline function of degree three using Scipy \cite{2020SciPy-NMeth} and calculate the fit coefficients.
Finally, we can calculate with these coefficients the values of $J$ and ${\delta J}/{\delta V_{k}}$ 
at new values of $V_k \in \{V_s, V_\ell\}$ during the optimization. 
To obtain a good fit, we should have enough pre-stored data set of triples $(V_s, V_\ell, J)$, 
and we see that a grid of $100\times 100$ is sufficient.
As explained in 
Appendix~\ref{sec:gradient:methods}, the spline-fit method is computationally 
efficient and enables faster optimization compared to calculating 
$J$ and ${\delta J}/{\delta V_{k}}$ analytically or 
with the finite-difference method using Eq.~\eqref{v_to_j}.  

\begin{figure}
\includegraphics{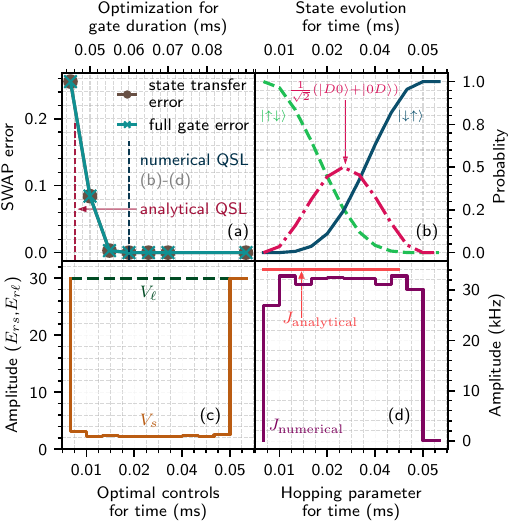}
\caption{
(a) The error of the SWAP gate is numerically minimized for different gate durations
based on a Fermi-Hubbard model; 
the numerical quantum speed limit~(QSL) of $0.06$~ms is close
to the analytical one from Sec.~\ref{analytical} with $0.046$~ms.
The full gate error defined by Eq.~\eqref{gate_cost_function} matches the state transfer error defined by Eq.~\eqref{cost_function}.
(b)~Corresponding state evolution of duration $0.06$~ms
from $\ket{\uparrow \downarrow}$
to $ \ket{\downarrow \uparrow}$ with an intermediate state 
$(\ket{D0}{+}\ket{0D})/\sqrt{2}$. The symmetry in the Hamiltonian of Eq.~\eqref{eqn_for_motion_mat} also suggests that the same pulse
will transfer the state $\ket{\da\ua}$ to $\ket{\ua\da}$.
(c)-(d)~Time dependence of the optimized 
lattice parameters $V_s$ and $V_\ell$ and
hopping parameter $J_\mathrm{numerical}$;
a minimum $V_s$ results in a maximal
$J_\mathrm{numerical}$ having a similar form as $J_\mathrm{analytical}$.
The discrepancy between $J_\mathrm{numerical}$ and $J_\mathrm{analytical}$ results from 
the constraints included in the optimization.
\label{two_band_opt}}
\end{figure}

We begin by optimizing the SWAP gate across multiple gate durations, 
ranging from $0.03$ ms to $0.09$ ms and each time step $\Delta t=0.005$ ms. For a simple optimization example, 
we use a linearly decreasing $V_s$ as an initial guess for the 
pulse sequence while keeping $V_\ell = 30$$E_{r\ell}$ constant. 
The bounds for $V_s$ are fixed between $2$$E_{rs}$ and $30$$E_{rs}$.
The scattering length $a$ is set to zero for the SWAP gate.
The numerical optimization imposes constraints on the start and the end 
of the pulses to ensure that it is experimentally feasible. These constraints, 
in turn, force the resulting hopping parameter $J_{\mathrm{numerical}}$ to start and end at zero. 
Figure~\ref{two_band_opt}(a) illustrates the optimized SWAP gate error 
as a function of the gate duration. 
The infidelity decreases 
rapidly, becoming less than $10^{-8}$ for durations longer than $0.06$ ms, 
which is close to the analytical quantum speed limit of $0.046$ ms, calculated in the Appendix.~\ref{app:analytic}.
However, these 
additional constraints prevent the optimization from fully achieving the analytical quantum speed limit.
Now, as explained in Sec.~\ref{objective}, for the two-band Fermi-Hubbard model, basis states $\ket{\ua\ua}$ and $\ket{\da\da}$
do not change under SWAP, and from the symmetry of the Hamiltonian in Eq.~\eqref{eqn_for_motion_mat}, we must reach the state $\ket{\ua\da}$ from the initial state $\ket{\da\ua}$. We verify this explicitly by calculating the full SWAP gate error as
\begin{equation}
C_{SWAP}^{F}=1-\frac{1}{N_i}\sum_{(\Psi_\text{ini},\Psi_\text{tar})}
\hspace{-2mm}\abst{\langle \Psi_\text{tar} \vert\, \mathcal{U}_{N_T}\!\cdots \mathcal{U}_{1}  \vert \Psi_\text{ini} \rangle}^2,
 \label{gate_cost_function}
\end{equation}
where $(\Psi_\text{ini},\Psi_\text{tar})$ denotes the $N_i$ different possible tuples of initial and target states. For the SWAP gate with the two-band Fermi Hubbard model, we take two sets $\{\ket{\ua\da},\ket{\da\ua}\}$ and $\{\ket{\da\ua},\ket{\ua\da}\}$, an calculate the full SWAP gate error.
We show that the full SWAP gate error matches exactly with the error of transferring the state from $\ket{\ua\da}$ to $\ket{\da\ua}$.
The evolution of the three basis states is depicted in Figure~\ref{two_band_opt}(b).
Starting from $\ket{\ua\da}$, the system evolves into $\ket{\da\ua}$, 
while the probability of the third state $(\ket{D0}+\ket{0D})/\sqrt{2}$ peaks at half of the gate duration.
The symmetry in the time evolution also suggests that the same pulse
will transfer the state $\ket{\da\ua}$ to $\ket{\ua\da}$.
The optimized pulses $V_s$ and $V_\ell$ for the $0.06$ ms gate 
duration are shown in Figure~\ref{two_band_opt}(c) where $V_s$ approaches the 
minimum bound of $2$$E_{rs}$. 
Aside from the additional imposed bounds, $J_{\mathrm{numerical}}$ 
attempts to reach the maximum $J=34.03$ kHz and closely resembles $J_{\mathrm{analytical}}$ as shown in Figure~\ref{two_band_opt}(d).
\begin{res}\label{result-two}
The shortest, numerically optimized SWAP gate has a duration of $0.06$~ms with an upper bound of
$J_{max}=34.03$~kHz. The corresponding pulse resembles the analytical time-optimal pulse except for
the additional constraints imposed in the numerical calculations, which forces $J_{\mathrm{numerical}}$ to start and end at zero.
\end{res}

\section{Effect of higher bands and offsite terms\label{Excitation to higher bands}}
\subsection{Hamiltonian description\label{hamiltonian_des}}
As discussed in the previous section, atoms in the double well
are described by the two-band Fermi-Hubbard model and
characterized by the nearest-neighbor hopping $J$ and the onsite interaction $U$. 
However, if the pulses are changed non-adiabatically, the atoms 
may be excited within the lattice, causing the two-band Fermi-Hubbard model to 
fail in accurately describing the system. In such cases, it is necessary 
to account for the higher bands of the Fermi-Hubbard model and the offsite interactions between atoms located on different sites in the double well.
To incorporate these higher bands and offsite interactions, we generalize 
the two-band Hamiltonian as shown in Fig.~\ref{fig:schematic_3}. 
The bands appear in pairs, with each pair corresponding to one level $p$ of the Fermi-Hubbard model 
so we have $M{=}b/2$ levels for $b$ bands. 
For instance, four bands form two levels with $p\in\{0,1\}$, where the zeroth $(b=0)$ band and 
the first  ($b=1$) band form the level $p=0$, and the second ($b=2$) and the third ($b=3$) band form the level $p=1$.

\begin{figure}
    \includegraphics{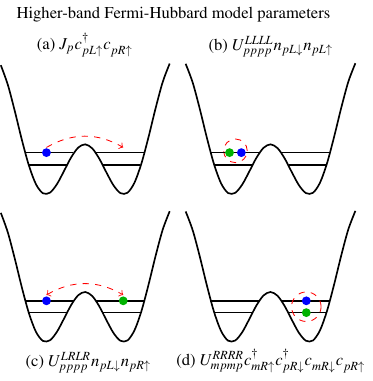}
    \caption{Different hopping and interaction processes in a symmetric double well
    for a Fermi-Hubbard model
    with higher bands. Each level $p$ in the double well is made from two energy bands as shown in Fig.~\ref{fig:schematic_3}.
    (a)~Single-atom hopping term [see~Eq.~\eqref{hopping}] between wells for level $p$ with hopping strength $J_p$ 
    calculated from Eq.~\eqref{v_to_j_general}.
    (b)-(d)~Two atoms of opposite spin interact 
    with strength 
    $U^{\alpha\beta\gamma\delta}_{mnop}$ calculated from Eq.~\eqref{eqn_for_U_general}, where $\alpha,\beta,\gamma,\delta $ refer to the left (L) or the right (R) well 
     and $m,n,o,p$ denote different levels:  
     (b)~$U_{pppp}^{LLLL}$ has two atoms on level $p$ in the left (L) well [see~Eq.~\eqref{fermi_hubbard_2}] ;
    (c)~$U_{pppp}^{LRLR}$ has two atoms on level $p$ in different wells [see~Eq.~\eqref{fermi_hubbard_3}]; 
    (d)~$U_{mpmp}^{RRRR}$ has two atoms on different levels $p,m$ but in the right (R) well [see~Eq.~\eqref{fermi_hubbard_4}].
    \label{fig:schematic}}
\end{figure}

The extended higher-band Fermi-Hubbard Hamiltonian for a double well is given by  
\begin{subequations}\label{fermi_hubbard_higher}
\begin{align}
\hat{H}=&-\sum_{p}\sum_{\sigma}J_p(c_{pL\sigma}^{\dagger}c^{\phantom{\dagger}}_{pR\sigma}{+}h.c.) \label{hopping}\\
&+\sum_{m,n,o,p}\;\sum_{\alpha,\beta,\gamma,\delta}U^{\alpha\beta\gamma\delta}_{mnop}c_{m\alpha \ua}^\dagger c_{n\beta \da}^\dagger
c^{\phantom{\dagger}}_{o\gamma\da}c^{\phantom{\dagger}}_{p\delta\ua}\label{interaction_term}\\
&+ \sum_{p}\sum_{\alpha}\sum_{\sigma}\epsilon_{p\alpha}\mathfrak{n}_{p\alpha\sigma} \label{chemical_potential},
\end{align}
\end{subequations}
where $\alpha,\beta,\gamma,\delta \in \{L,R\}$, $m,n,o,p \in \{1,\ldots, M\}$ and $\sigma \in \{\ua,\da\}$. 
The term in Eq.~\eqref{hopping} describes hopping with amplitudes $J_p$ for different levels $p$ of the well, as illustrated in Fig.~\ref{fig:schematic}(a). Specifically, $J_0$ corresponds to the hopping term $J$ in the two-band model. The hopping amplitude $J_p$ is given by
\begin{equation}\label{v_to_j_general}
J_p=-\int w_{pL}(x) \left[-\frac{\hbar^{2}}{2\mass}\partial_{x}^{2}+V(x)\right] w_{pR}(x) \, dx.
\end{equation}
The interaction strength $U_{mnop}^{\alpha\beta\gamma\delta}$ is calculated using the Wannier functions as follows:
\begin{equation}\label{eqn_for_U_general}
U_{mnop}^{\alpha\beta\gamma\delta}\!=\!\!\iint \! U_{3 \mathrm{D}} w_{m\alpha}^\dagger(\mathbf{r}_1) w^\dagger_{n\beta}(\mathbf{r}_2) w_{o\gamma}(\mathbf{r}_2) w_{p\delta}(\mathbf{r}_1)d\mathbf{r}_1  d\mathbf{r}_2,
\end{equation}
where $U_{3 \mathrm{D}}{:=}U_{3 \mathrm{D}}(\mathbf{r}_1,\mathbf{r}_2)$.
Finally, the Hamiltonian includes an onsite energy term corresponding to the 
energies $\epsilon_{pj}$ in Eq.~\eqref{chemical_potential}, which are calculated from 
\begin{equation}\label{onsite energy}
\epsilon_{p\alpha}=\int w_{p\alpha}(x) \left[-\frac{\hbar^{2}}{2\mass}\partial_{x}^{2}+V(x,t)\right] w_{p\alpha}(x) \, dx.
\end{equation}
While the Wannier functions associated with different bands are orthonormal, resulting in zero overlap, the integral
over four different Wannier functions used in Eq.~\eqref{eqn_for_U_general} is not necessarily zero. However, the interaction strength \( U_{mnop}^{\alpha\beta\gamma\delta} \), which is determined by the Wannier functions, is typically negligible when \( m \neq n \neq o \neq p \).
Thus we consider just
the following interaction terms from Eq.~\eqref{interaction_term} with significant contributions:
\begin{subequations}
\begin{align}
&\hat{H}_{\text{int}}=\sum_{p}\sum_{\alpha}U_{pppp}^{\alpha\alpha\alpha\alpha}\mathfrak{n}_{p\alpha\da}\mathfrak{n}_{p\alpha\ua} \label{fermi_hubbard_2}\\ 
&+\sum_{p,m}\sum_{\alpha,\beta}\sum_{\sigma} U_{mppm}^{\beta\alpha\alpha\beta}\mathfrak{n}_{p\alpha\sigma}\mathfrak{n}_{m\beta(-\sigma)} \label{fermi_hubbard_3}\\ 
&+\sum_{p,m}\sum_{\alpha,\beta} U_{mpmp}^{\beta\alpha\alpha\beta}c_{m\beta\ua}^\dagger c_{p\alpha\da}^\dagger 
c^{\phantom{\dagger}}_{m\beta\da}c^{\phantom{\dagger}}_{p\alpha\ua} \label{fermi_hubbard_4}\\
&+\sum_{p,m}\sum_{\alpha,\beta} U_{ppmm}^{\alpha\alpha\beta\beta}c_{p\alpha\ua}^\dagger c_{p\alpha\da}^\dagger 
c^{\phantom{\dagger}}_{m\beta\da}c^{\phantom{\dagger}}_{m\beta\ua} \label{fermi_hubbard_5}\\
&+\sum_p\sum_{\alpha}\sum_{\sigma}U_{pppp}^{\alpha\alpha L R}\, \mathfrak{n}_{p\alpha,-\sigma}\, (c_{pL\sigma}^{\dagger}c^{\phantom{\dagger}}_{pR\sigma}{+}h.c.) \label{hopping_correction}.
\end{align}
\end{subequations}
\begin{rem}
For the Eqs.~\eqref{fermi_hubbard_3}-\eqref{fermi_hubbard_5}, we only consider interaction terms such that the two atoms either have the same energy levels $p$ and $m$ or they sit on the same side $\alpha$ and $\beta$ of the double well, i.e.,
\begin{equation*}
p=m \,\text{ if }\, \alpha\neq \beta
\;\text{ and }\; \alpha=\beta \,\text{ if }\, p\neq m.
\end{equation*}
\end{rem}

The interaction term in Eq.~\eqref{fermi_hubbard_2} involves onsite interactions with strengths $U_{pppp}^{\alpha\alpha\alpha\alpha}$, corresponding to interactions at level $p$ and side $\alpha$, as shown in Fig.~\ref{fig:schematic}(b). Additionally, during gate operations, the short lattice depth $V_s$ decreases significantly, as depicted in Fig.~\ref{two_band_opt}(c). Consequently, the barrier between the double well becomes sufficiently small that atoms on either side or at different levels start to interact. This is represented by the offsite interaction term $\mathfrak{n}_{p\alpha\sigma}\mathfrak{n}_{m\beta(-\sigma)}$, proportional to $U_{mppm}^{\beta\alpha\alpha\beta}$, as described in Eq.~\eqref{fermi_hubbard_3} and illustrated in Fig.~\ref{fig:schematic}(c).
The term $ c_{m\beta\ua}^\dagger c_{p\alpha\da}^\dagger  c^{\phantom{\dagger}}_{m\beta\da}c^{\phantom{\dagger}}_{p\alpha\ua}$ in Eq.~\eqref{fermi_hubbard_4} and Fig.~\ref{fig:schematic}(d) describes the spin exchange process with strength $U_{mpmp}^{\beta\alpha\beta\alpha}$ between two atoms at levels $p$ and $m$ and sides $\alpha$ and $\beta$. The term $c_{p\alpha\ua}^\dagger c_{p\alpha\da}^\dagger c^{\phantom{\dagger}}_{m\beta\da}c^{\phantom{\dagger}}_{m\beta\ua} $ in Eq.~\eqref{fermi_hubbard_5} represents the correlated pair tunneling of two atoms within the double well.

Lastly, we have a correction term proportional to $\Delta J_{p\alpha}=U_{pppp}^{\alpha\alpha L R}$ in Eq.~\eqref{hopping_correction} yields
\begin{equation}\label{eqn_for_delta_j}
\Delta J_{p\alpha}=  \iint U_{3 \mathrm{D}}\, |w_{p\alpha}(\mathbf{r}_1)|^2 w_{pL}(\mathbf{r}_2) w_{pR}(\mathbf{r}_1)d\mathbf{r}_1  d\mathbf{r}_2.
\end{equation}
This term accounts for density-assisted hopping, which corrects the hopping parameter $J_p$ when another atom of opposite spin is in the double well on side $\alpha$.

\subsection{Simulations}\label{sim_higher_band}
To study the effects of these new terms, we perform a time evolution of 
the system using the Hamiltonian in Eq.~\eqref{fermi_hubbard_higher} 
with a non-adiabatic approach. Because the $x$ direction terms of the potential in Eq.~\eqref{potential_2} change non-adiabatically while keeping the $y$ and $z$ direction terms constant, 
the Wannier functions in the $x$ direction vary at each time step during the evolution. 
Consequently, after each unitary evolution at time step $t$, we compute the updated 
Wannier functions $w_t$, and transform the evolved state onto these new Wannier states.
The non-adiabatic time evolution is described by
\begin{equation}\label{time_evol_proj}
\Psi(t{+}1)=P(w_{t+1},w_t)\, e^{-iH_t\delta t}\Psi(t).
\end{equation}
Here, $P(w_{t+1},w_t)$ denotes the basis transformation operator, which describes the non-adiabaticity of the time evolution by changing the basis of the state into the new Wannier basis states at every time step.
Since calculating the Wannier functions as described in Sec.~\ref{Two-band-Fermi-Hubbard} is 
necessary to construct $P(w_{t+1},w_{t})$ for each time step, 
we cannot utilize the spline method employed in Sec.~\ref{numeric}, and
instead, we must directly compute the parameters $J$, $U$, $\Delta J$, and $\epsilon$ 
from Eqs.~\eqref{v_to_j_general}-\eqref{onsite energy} for each time step.

\begin{figure}
\includegraphics{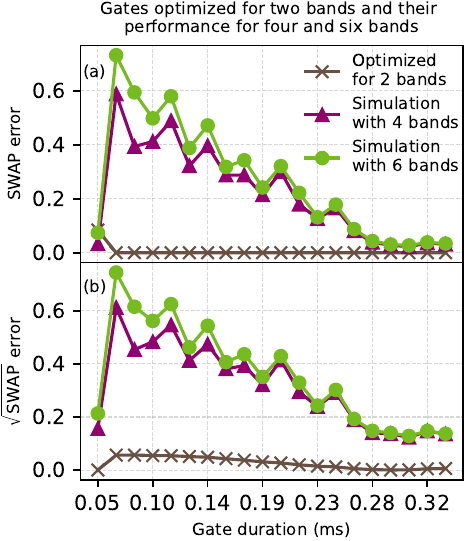}
\caption{The SWAP and $\sqrt{\mathrm{SWAP}}$ gates have been numerically optimized
using a two-band Fermi-Hubbard model for different gate durations as in Fig.~\ref{two_band_opt}(a).
Their performance is compared to Fermi-Hubbard simulations with four and six bands.
(a)~For SWAP, we see
much higher gate errors for four and six bands
as compared to two bands
for gate durations of less than $0.28$~ms while
(b)~high gate errors are observed for $\sqrt{\mathrm{SWAP}}$
with four and six bands and all gate durations.}
\label{fig:two_vs_higher}
\end{figure}

We use this new simulation method to assess the validity of the two-band model. We simulate the four-band and six-band Fermi-Hubbard model using the pulses optimized from the two-band model. We optimize the SWAP gate for various durations ranging from $0.05$ ms to $0.35$ ms within the two-band Fermi-Hubbard model [see Sec.~\ref{numeric}]. Figure~\ref{fig:two_vs_higher}(a) presents the infidelities for the two-band model at different gate durations. It is evident that high-fidelity gates can be achieved for very short gate durations using the two-band model. We then apply these optimized control pulses to simulate the four-band and six-band models. As shown in Fig.~\ref{fig:two_vs_higher}(a), the simulations with higher bands exhibit significant deviations from the two-band results, with the discrepancies increasing for shorter gate durations. 

We conduct a similar analysis for the $\sqrt{\mathrm{SWAP}}$ gate, as shown in Fig.~\ref{fig:two_vs_higher}(b). Here, we use the optimized time-dependent lattice depths $V_s$ and $V_\ell$ from the SWAP gate optimizations and optimize only the time-independent scattering length $a$ using the two-band model. The interaction strength $U$ is calculated from $a$ using Eq.~\eqref{eqn_for_U}. Figure~\ref{fig:two_vs_higher}(b) shows that, similar to the SWAP gate, very low infidelities for the $\sqrt{\mathrm{SWAP}}$ gate can be achieved using the two-band Fermi-Hubbard model. 
Next, we simulate the four-band and six-band models using the optimal controls from the two-band model. For short gate durations, we again observe significant divergence of the higher-band simulations from the two-band model results [see Fig.~\ref{fig:two_vs_higher}(b)]. However, unlike the SWAP gate, the higher-band simulations fail to match the two-band results even for longer gate durations in the case of the $\sqrt{\mathrm{SWAP}}$ gate. This discrepancy is likely due to the additional off-site interactions present in the higher-band model [see Eq.~\eqref{fermi_hubbard_higher}], whereas the two-band model includes only onsite interactions.
Furthermore, in the two-band model, the Wannier states used in Eq.~\eqref{eqn_for_U} are computed at the start of the time evolution and are assumed to remain constant throughout. This assumption holds for an adiabatic time evolution, where the initially calculated Wannier states are eigenstates of the Hamiltonian at all times. However, for fast gates with non-adiabatic changes in lattice depths, the Wannier states $w_{pL}(t)$ and $w_{pR}(t)$ vary during the evolution. This variation introduces additional errors in the gate fidelity, which are accounted for in the higher-band simulations through the basis transformation operators in Eq.~\eqref{time_evol_proj}. 

\begin{algorithm}[t]
\caption{Optimization of the SWAP and $\sqrt{\mathrm{SWAP}}$ gates with higher-band models}\label{optimization_algorithm_higher_band}
\begin{algorithmic}
\State \textbf{SWAP}:
Same as in the SWAP optimization in Algorithm~\ref{optimization_algorithm_two_band}.
Instead of the spline-fit method,\\ a combination of approximate analytical and finite-difference gradients is used.
\vspace{0.3cm}
\State $\mathbf{\sqrt{\mathrm{\mathbf{SWAP}}}}$:
\State \textbf{Optimization parameters}: $V_s(t), V_\ell (t)$, and $a$
\State \textbf{Step 1}: Define the $\sqrt{\mathrm{SWAP}}$-gate cost function $C_{\sqrt{\mathrm{SWAP}}}=1- \abst{\langle {\Psi_{\sqrt{\mathrm{SWAP}}} \vert \Psi(T)}\rangle}^2$
\State \textbf{Step 2}: Minimize $C_{\sqrt{\mathrm{SWAP}}}$ and optimize $V_s(t)$ and $V_\ell(t)$ using a gradient-based optimization
\State \textbf{Step 3}: Perform a one-dimensional search to find the optimized $a$ for $C_{\sqrt{\mathrm{SWAP}}}$
\State \textbf{Output}: Optimized $V_s(t), V_\ell (t)$, and $a$ for $\sqrt{\mathrm{SWAP}}$
\end{algorithmic}
\end{algorithm}

\begin{res}\label{result-three}
Simulations in higher-band Fermi-Hubbard models of pulses optimize using a two-band model show significantly higher errors for the SWAP and $\sqrt{\mathrm{SWAP}}$ gates. This suggests that for non-adiabatic gate operations, the two-band model is not sufficient and optimizations using higher-band Fermi-Hubbard models are essential.  
\end{res}
\section{Optimization with higher-band Fermi-Hubbard model\label{optimization with higher band}}
We showed in Sec.~\ref{Excitation to higher bands} that the optimization with a two-band Fermi-Hubbard model is 
insufficient in describing the fast SWAP and $\sqrt{\mathrm{SWAP}}$ 
gates. To better capture the behavior of the system and generate efficient gates,
we must include higher bands of the model into our optimization.
In analogy to the two-band model, we optimize the SWAP and $\sqrt{\mathrm{SWAP}}$ 
gates using gradient-based methods with a higher-band Fermi-Hubbard model. 
We use our non-adiabatic simulation method described in Sec.~\ref{sim_higher_band}.
We optimize $V_s$ and $V_\ell$ for the SWAP gate with 
the scattering length $a=0$. In contrast to the two-band model,
where the $\sqrt{\mathrm{SWAP}}$ gate is controlled by the lattice depths $V_s$ and $V_\ell$ from the SWAP gate optimization, we independently optimize $V_s$ and $V_\ell$ for the target state $\Psi_{\sqrt{\mathrm{SWAP}}}$ and then perform a one-dimensional search to find the optimal $a$. This is described in the Algorithm~\ref{optimization_algorithm_higher_band}.
For the two-band model in Sec.~\ref{numeric}, we calculate the gradient of
the cost function $C$ with respect to $V_s$ and $V_\ell$ using spline-fit method
whereas the gradient for the one-dimensional search is trivial and easily computed using finite-differences. 
However for the higher-band model, in the absence of a spline-fitting approach, the gradients ${d C}/{d V_{s}}$ 
and ${d C}/{d V_{\ell}}$ 
can be computed either 
analytically or through finite-differences. In Appendix~\ref{approximate-analytic-gradient}, we derive an approximate analytical expression for the gradient that is considerably faster than the finite-difference method as shown in Appendix~\ref{sec:gradient:methods}. 
This approximation allows for an accelerated optimization process. 
Subsequently, we can employ the finite-difference method to refine the optimization, 
enabling convergence to the minimum of the cost function.

\subsection{State-to-state transfer optimization}\label{higher_band_results}
\begin{figure}
\includegraphics{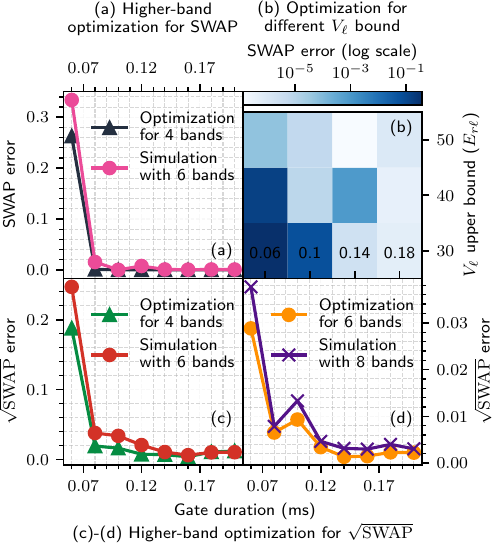}
\caption{Higher-band Fermi-Hubbard optimizations for
SWAP and $\sqrt{\mathrm{SWAP}}$.
(a)~SWAP gate optimizations for four bands have an error of less than $0.001$
for gate durations of more than $0.08$~ms.
These optimized pulses for four bands are used
in six-band simulations which match closely with errors of less than $0.001$ 
for gate durations larger than $0.1$~ms. This suggests that four-band optimizations
are sufficient for the high-fidelity SWAP gate.
(b)~SWAP gate optimization errors for upper bounds
given by 30, 40, and 50$E_{r\ell}$
for $V_\ell$ and the four-band model;
For each upper bound, the error is shown with gate durations of $0.06$, $0.1$, $0.14$, and $0.18$~ms. 
Larger upper bounds result in smaller gate errors, especially for shorter gate durations. 
(c)~Similar optimizations and simulations for $\sqrt{\mathrm{SWAP}}$.
The four-band optimizations have errors of less than $0.007$ for gate durations larger than $0.12$~ms.
The corresponding six-band simulations slightly differ with errors of less than $0.007$ 
for gate durations larger than $0.16$~ms. 
(d)~Improved $\sqrt{\mathrm{SWAP}}$ optimizations
with errors of less than $0.005$ for gate durations larger than
$0.12$~ms
by increasing the upper bound for $V_\ell$  to $45$~$E_{r\ell}$ and optimizing
for six bands. The corresponding eight-band simulations 
have errors of less than $0.005$ for gate durations larger than $0.12$~ms
and show negligible excitation beyond six bands.  
\label{fig:higher_band_opt}}
\end{figure}

First, we optimize the transfer from the initial state $\Psi_0$ to the target state $\Psi_{\mathrm{SWAP}}$,
for different gate durations using the 
four-band Fermi-Hubbard model. We try to find the optimal $V_s$ and $V_\ell$ for variable gate durations from $0.06$ ms to $0.20$ ms. 
We constrain the optimization with bounds on $V_s$ and $V_\ell$ given by the pairs $(0.1$~$E_{rs}$, $45$~$E_{rs})$ and $(7$~$E_{r\ell},$ $35$~$E_{r\ell})$ respectively.
The optimized SWAP infidelities are presented in Fig.~\ref{fig:higher_band_opt}(a) where the error 
is less than $0.001$ for gate durations larger than $0.08$~ms. 
We emphasize that compared to the four-band simulation in Fig.~\ref{fig:two_vs_higher}(a), 
we achieve a significantly lower infidelity after optimizing for the four-band 
model with similar gate durations. 
The improvement in fidelity comes from the enhanced controllability by including the higher bands 
with multiple hopping parameters.
To validate the four-band model, we simulate the system with 
six bands using the optimized controls from the four-band optimization. 
The gate error in the six-band model shows negligible deviation from the 
error in the four-band model [see Fig.~\ref{fig:higher_band_opt}(a)]. 
Specifically, the error in the six-band simulation is less than 0.001 for 
gate durations greater than 0.1 ms. This result suggests that excitations 
beyond four bands are negligible, indicating that the four-band Fermi-Hubbard model is 
sufficient to capture most the dynamics of the double-well system under the SWAP gate.
We also examine the effect of the upper bound of the long lattice depth $V_\ell$ on the gate fidelity. 
The SWAP gate is optimized for upper bounds given by 30, 40, and 50 times the value of $E_{r\ell}$, 
and the resulting infidelities are shown in Fig.~\ref{fig:higher_band_opt}(b) on a logarithmic scale. 
We observe that gate fidelity improves with a higher upper bound on $V_\ell$, as increased 
$V_\ell$ leads to more localized Wannier states and enhances the hopping strength $J$.
\begin{figure*}  
 \includegraphics{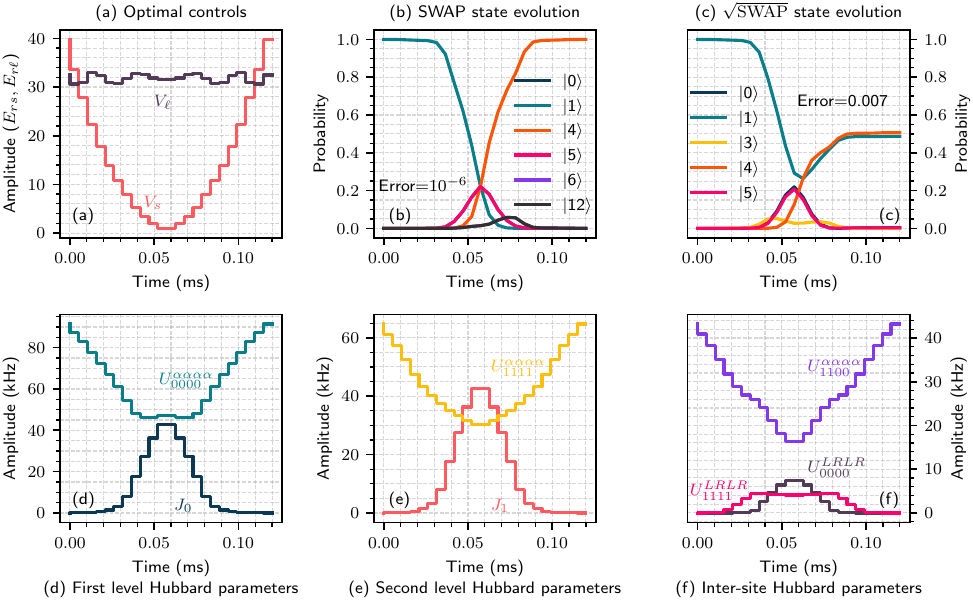}
 \caption{Four-band Fermi-Hubbard model optimizations for a short gate duration of $0.12$~ms
 relying on excitations to higher bands
 and multiple hopping and interaction strengths.   
(a)~Optimal controls $V_s$ and $V_\ell$ in their respective recoil energies 
$E_{r{s(\ell)}}=\hbar^2k_{s(\ell)}^2/(2m)$. 
(b)~State evolution under SWAP from $ \ket{1}=\ket{ \uparrow \downarrow}$ to
$ \ket{4}=\ket{\downarrow \uparrow}$ for $0.12$~ms  
with optimal controls from~(a) and 
state labeling scheme as in Sec.~\ref{three_four_atom} and Table.~\ref{table:table_state}.
(c)~State evolution under $\sqrt{\mathrm{SWAP}}$ from $ \ket{1}=\ket{\uparrow \downarrow}$
to $[{(1{+}i)}\ket{1}{-} {(1{-}i)}\ket{4}]/2=[{(1{+}i)}\ket{\ua\da} {-} {(1{-}i)}\ket{\da\ua}]/2$ with optimal controls from (a) and optimized scattering length $a=1995.22 \times a_0$ where
$a_0=5.29\times10^{-11}$~m is the Bohr radius.
(d)-(f)~The corresponding Fermi-Hubbard parameters: Each level $p$ in the double well consists of two energy bands $2p$ and $2p{+}1$,
resulting in two levels for the four-band Fermi-Hubbard model.
(d)~The hopping parameter {$J_0$} on the first level and between the left 
and the right side of the double well increases with decreasing $V_s$ and 
increasing $V_\ell$. The onsite interaction $U^{\alpha\alpha\alpha\alpha}_{{0000}}$ on the first level
is directly proportional to $V_s$ and $V_\ell$. 
(e)~The second-level hopping parameter $J_1$ and the
onsite interaction $U^{\alpha\alpha\alpha\alpha}_{{1111}}$ behave similar as in (d).
(f)~The offsite interaction $U^{LRLR}_{{0000}}$ and $U^{LRLR}_{{1111}}$ 
 between the left and right sides on the first and the second level 
 have a small magnitude and become significant for 
 small $V_s$. Here, $U^{\alpha\alpha\alpha\alpha}_{{1100}}$ represents the interaction 
 between the first and the second level on the same side of the double well and behaves similar as $U^{\alpha\alpha\alpha\alpha}_{pppp}$.
  \label{fig:v_to_hubbard}}
\end{figure*}

Next, we optimize the lattice depths $V_s$ and $V_\ell$ and the scattering length $a$ for the 
$\sqrt{\mathrm{SWAP}}$ gate i.e., for transferring the state $\Psi_0$ to the state $\Psi_{\sqrt{\mathrm{SWAP}}}$, using the same initial conditions and bounds as in the 
SWAP-gate optimization. As shown in Fig.~\ref{fig:higher_band_opt}(c), the 
infidelity increases for shorter pulse durations. The four-band optimizations 
yield errors of less than 0.007 for gate durations longer than 0.12 ms.
Similar to the SWAP gate, we compare our results with six-band simulations, 
as illustrated in Fig.~\ref{fig:higher_band_opt}(c). The six-band simulations differ 
slightly from the four-band results, showing errors of less than 0.007 for gate 
durations longer than 0.16 ms, indicating the presence of higher-band excitations in the system. 
This can be mitigated by increasing the upper bound on $V_\ell$ to 45~$E_{r\ell}$ 
and optimizing within the six-band model, as shown in Fig.~\ref{fig:higher_band_opt}(d).
We further compare the optimization in the six-band model with simulations in the eight-band 
model and observe negligible excitations beyond six bands.
Both the six-band optimization and the eight-band simulation result in gate errors smaller 
than 0.005 for gate durations longer than 0.12 ms.
This suggests that for the $\sqrt{\mathrm{SWAP}}$ gate, good pulses can be identified 
through four-band or six-band optimizations where the six-band model gives lower 
higher-band excitation compared to the four-band model. 

\subsection{Dynamics with optimzed controls}
We focus on one set of optimal lattice depths $V_{s}$ and $V_{\ell}$ for a duration of 0.12 ms
as shown in Fig.~\ref{fig:v_to_hubbard}(a), which have been optimized for the $\sqrt{\mathrm{SWAP}}$ gate. The optimal scattering length is $a = 1995.22~a_0$, where $a_0 = 5.29 \times 10^{-11}$ m is the Bohr radius. The controls are simple and realistic, adhering to experimental constraints. The corresponding time evolution for the SWAP and $\sqrt{\mathrm{SWAP}}$ gates is illustrated in Fig.~\ref{fig:v_to_hubbard}(b)-(c). The system begins in the state $\Psi_0$ and evolves to the target states $\Psi_{\mathrm{SWAP}}$ and $\Psi_{\sqrt{\mathrm{SWAP}}}$ for the SWAP and $\sqrt{\mathrm{SWAP}}$ gates, respectively. 
The symmetry in the system suggests that the same pulse
will transfer the state $\Psi_{\mathrm{SWAP}}$ to $\Psi_0$ for SWAP gate.
Notably, there is excitation and de-excitation from higher-band states during the time evolution, resulting in faster gate operations.
The combined probabilities for all the states involving the second level are less than $10^{-7}$ and $10^{-4}$ at the end of the \text{SWAP} and $\sqrt{\text{SWAP}}$ gates, respectively. This suggests that the optimized controls effectively minimize excitations to higher levels at the end of the gates, which is crucial for any quantum operation. The remaining error arises from probabilities on the order of $10^{-3}$ for the first-level states $\ket{D0}$ and $\ket{0D}$ in the case of the $\sqrt{\text{SWAP}}$ gate.

We also examine the impact of the optimal $V_s$, $V_\ell$, and $a$ on the higher-band Fermi-Hubbard parameters of Eq.~\eqref{fermi_hubbard_higher}. Each level $p$ in the double well is composed of two energy bands, {$2p$ and $2p+1$}, yielding two levels in the four-band Fermi-Hubbard model. The hopping parameter {$J_0$} represents the hopping between the left (L) and right (R) sides of the well on the first level (or equivalently the hopping $J$ of the two-band Fermi-Hubbard model)
and $U^{\alpha\alpha\alpha\alpha}_{{0000}}$ corresponds to the interaction $U$ of the two-band Fermi-Hubbard model as shown
in Fig.~\ref{fig:v_to_hubbard}(d). The value of $J_0$ increases with decreasing $V_s$ and increasing $V_\ell$, as these conditions lead to a higher overlap between the Wannier functions $w_{{0}L}$ and $w_{{0}R}$. Conversely, the onsite interaction $U^{\alpha\alpha\alpha\alpha}_{{0000}}$ decreases with decreasing $V_s$ as shown in Fig.~\ref{fig:v_to_hubbard}(d). 

For the two-band Fermi-Hubbard model optimization in Sec.~\ref{numeric}, we assume that the interaction strength is independent of changes in the Wannier functions and remains constant and proportional to $a$. However, we observe that this assumption breaks down with our optimal lattice depths for the four-band Fermi-Hubbard model, as $V_s$ becomes shallow for fast gates.
In the four-band Fermi-Hubbard model, an additional hopping parameter {$J_1$} and onsite interaction strength $U^{\alpha\alpha\alpha\alpha}_{{1111}}$ arise, as shown in Fig.~\ref{fig:v_to_hubbard}(e). The parameter {$J_1$} has a larger magnitude compared to {$J_0$} because the overlap between the Wannier functions {$w_{1L}$} and {$w_{1R}$} for the second level is greater than that for the first level. Conversely, $U^{\alpha\alpha\alpha\alpha}_{{1111}}$ has a smaller magnitude compared to $U^{\alpha\alpha\alpha\alpha}_{{0000}}$ due to the reduced overlap of the Wannier functions on the same site at the second level.
Additionally, there are significant contributions from offsite interactions $U^{\alpha\beta\gamma\delta}_{mnop}$ when the lattice depths are shallow. 

Figure~\ref{fig:v_to_hubbard}(f) illustrates the variation of three different offsite interactions with time-dependent lattice depths. The terms $U^{LRLR}_{{0000}}$ and $U^{LRLR}_{{1111}}$ represent the offsite interactions between the left (L) and right (R) sides of the double well for the first and second levels, respectively. Similar to $J_p$, the offsite interaction $U^{LRLR}_{pppp}$ is proportional to the overlap between the wave functions on the left and right sides, thus it increases as $V_s$ decreases. The term $U^{\alpha\alpha\alpha\alpha}_{{1100}}$ represents the interaction between atoms occupying different levels but residing on the same side $\alpha$. For $U^{\alpha\alpha\alpha\alpha}_{pppp}$, the interaction $U^{\alpha\alpha\alpha\alpha}_{{1100}}$ decreases as the Wannier functions spread out with a shallower short lattice.

Note that the optimized gates are distinct from superexchange processes where $U \gg J$, which avoid the $\ket{D0}$ and $\ket{0D}$ states, but result in long gate times~\cite{doi:10.1126/science.1150841}. For the optimized pulses shown in Fig.~\ref{fig:v_to_hubbard}(a), we obtain $U^{\alpha\alpha\alpha\alpha}_{0000}/J_0 = 5.34$ and $U^{\alpha\alpha\alpha\alpha}_{1111}/J_1 = 3.64$, where these ratios are calculated by integrating the time-dependent curves in Fig.~\ref{fig:v_to_hubbard}(d)-(e). This ratio is close to the exchange ratio $U/J = 4/\sqrt{3}$, which permits population in the $\ket{D0}$ and $\ket{0D}$ states. 
Our optimized control operates near this exchange ratio, but is further refined to minimize population in the $\ket{D0}$ and $\ket{0D}$ states while achieving faster gates. Similar parameter regimes could be accessed through adiabatic control, but this would result in slower gates. The speedup in our case arises from non-adiabatically modulating $V_s$ and $V_l$ to dynamically enhance $J$ from an initially negligible value. This results in pulses that are experimentally feasible and helps manage excitations to higher energy levels efficiently.

\begin{res}\label{result-four}
Efficient SWAP and $\sqrt{\mathrm{SWAP}}$ gates are found
using higher-band Fermi-Hubbard models. The control duration can be as short as $0.08$~ms for transferring the state $\Psi_0$ to 
$\Psi_\mathrm{SWAP}$ and $0.12$~ms for transferring $\Psi_0$ to $\Psi_{\sqrt{\mathrm{SWAP}}}$ which is five times shorter than typical experimental state transfer durations \cite{Impertro_2024, chalopin2024opticalsuperlatticeengineeringhubbard}. The results for $\sqrt{\mathrm{SWAP}}$ are improved by optimizing with the six-band model and increasing the upper bound on $V_\ell$.
\end{res}

In the following Sections~\ref{three_four_atom} and
\ref{subsec:robustness}, we discuss how the obtained gates perform when they are applied to
error states with three and four atoms in a double well as well as
their robustness under multiple error sources. Afterwards, we consider the
full gate optimization using higher bands in the Fermi-Hubbard model (see Sec.~\ref{sec:full:gate}).

\section{Multi-atom dynamics \label{three_four_atom}}
We have demonstrated performance enhancements for SWAP and $\sqrt{\mathrm{SWAP}}$ for two atoms of opposite spins in a double well. However, in a real experimental setup, multiple double wells are controlled by global lasers $V_s$ and $V_\ell$, and some double wells may contain more or fewer than two atoms after state preparation. To fully characterize the system, it is essential to consider all possible atomic configurations within a double well and analyze the impact of our optimized pulses on these configurations.
For fermionic atoms, 16 different states can exist within a double well with up to four atoms:
\begin{itemize}
\item 0 atoms: $\ket{0 0}$

\item 1 atom: $\ket{\ua\!0}$, $\ket{0\!\ua}$, $\ket{\da\!0}$, $\ket{0\!\da}$
 
\item 2 atoms: $\ket{\ua\da}$, $ \ket{\da\ua}$, $\ket{\ua\ua}$, $\ket{\da\da}$, $\ket{D0}$, $\ket{0D}$

\item 3 atoms:
  $\ket{D\!\uparrow},  \ket{\uparrow\!D},  \ket{D\!\downarrow}, \ket{\downarrow\!D}$
  
\item 4 atoms: $\ket{DD}$,
\end{itemize}
where $ D = \uparrow \downarrow$ represents a double occupancy on one side of the double well. One-atom states serve as single-qubit states, and for two-qubit gates, states with three and four atoms are error states.
For adiabatic gates, the three-atom states behave similar to the one-atom states, and the four-atom state $\ket{DD}$ remains unchanged, as both sites are already fully occupied. However, for fast gate operations, these doubly occupied atoms can be individually excited to higher energy levels within the double well.

\begin{figure*}
    \centering
    \includegraphics{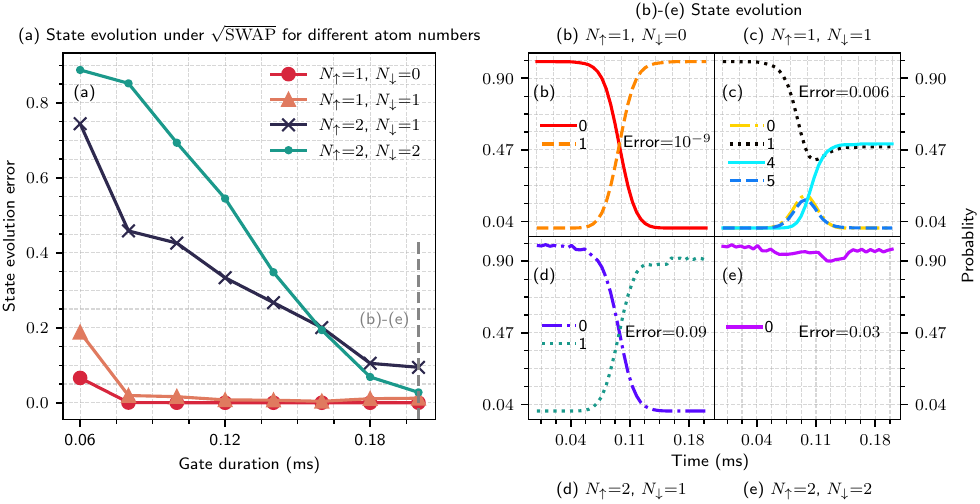}
    \caption{State evolution errors under $\sqrt{\mathrm{SWAP}}$ for different atom numbers using the optimized 
    control pulses from Fig.~\ref{fig:higher_band_opt}(c)
   as compared to the usual case of
   two atoms with $N_\uparrow{=}1$ and $N_\downarrow{=}1$ in a double well and measured with a four-band Fermi-Hubbard simulation.
(a)~Single atoms with $N_\uparrow{=}1$ and $N_\downarrow{=}0$ have an infidelity of less than $0.0005$ for gate durations larger than $0.08$~ms.
Two atoms with $N_\uparrow{=}1$ and $N_\downarrow{=}1$ agree with the optimized infidelities from Fig.~\ref{fig:higher_band_opt}(c).
The error states with three and four atoms with either
$N_\uparrow{=}2$ and $N_\downarrow{=}1$ or $N_\uparrow{=}2$ and $N_\downarrow{=}2$
result in higher infidelities
of around $10^{-2}$ for a duration of $0.20$~ms.
(b)-(e)~Corresponding state evolutions for a duration of $0.20$~ms. 
Further optimizations are possible for error states with three or four atoms if required.}
    \label{fig:3_4_atom}
\end{figure*}

We investigate the dynamics of various atomic configurations in a double well by using the method described in Appendix \ref{app:basis_labeling} for calculating and assigning the computational states with a given number of atoms. For this comparison, we employ the optimized controls $V_s$, $V_\ell$, and $a$ derived for the two-atom $\sqrt{\mathrm{SWAP}}$ gate from Sec.~\ref{higher_band_results}. The optimized gate errors for the two-atom case are repeated in Fig.~\ref{fig:3_4_atom}(a) for reference.
In the ideal scenario, the single-atom states undergo a SWAP operation from the left to the right side of the double well, or vice versa. This implies that our initial state for the one-atom case is $I = 0$ and the target state is $I = 1$. As illustrated in Fig.~\ref{fig:3_4_atom}(a), the state evolution error of the one-atom dynamics 
corresponding to a state with $N_\uparrow{=}1$ and $N_\downarrow{=}0$
is negligible, as the single atom is unaffected by the atom-atom interaction induced by $a$.
Similarly, in an ideal situation, the three-atom states should also perform a SWAP operation. For example, if the initial state is $\ket{D\!\ua}$, the target state should be $\ket{\ua\!D}$. However, as shown in Fig.~\ref{fig:3_4_atom}(a), the error for the three-atom configurations with $N_\uparrow{=}2$ and $N_\downarrow{=}1$ increase with shorter pulses due to the atom-atom interactions and excitations to higher energy levels.
A similar analysis was conducted for the four-atom state  with $N_\uparrow{=}2$ and $N_\downarrow{=}2$, where both the initial and target states are the same, i.e., $\ket{DD}$. This state exhibits higher state evolution error compared to the three-atom states, as more atoms are excited to higher levels and undergo multiple atom-atom interactions. The three- and four-atom configurations converge to infidelities below $0.1$ for a gate duration of $0.20$~ms.

The time evolution for one-, two-, three-, and four-atom cases under and a gate duration of $0.20$~ms is presented in Fig.~\ref{fig:3_4_atom}(b)-(e). In the one-atom case, the system evolves from the state $\ket{\ua\!0}$ to $\ket{0\!\ua}$, demonstrating the desired dynamics, as shown in Fig.~\ref{fig:3_4_atom}(b). For the two-atom case, the system performs the $\sqrt{\mathrm{SWAP}}$ gate with high fidelity, transitioning from $\Psi_0$ to $\Psi_{\sqrt{\mathrm{SWAP}}}$ as depicted in Fig.~\ref{fig:3_4_atom}(c). 
In the three-atom case, starting from $\ket{D\!\ua}$, the system evolves to $\ket{\ua\!D}$ with an infidelity of 0.09 [see Fig.~\ref{fig:3_4_atom}(d)]. Finally, the four-atom state $\ket{DD}$ is shown to excite to higher bands, resulting in an error of 0.027 in maintaining the $\ket{DD}$ state [see Fig.~\ref{fig:3_4_atom}(e)].
It is important to note that in Fig.~\ref{fig:3_4_atom}(b)-(e), we only display the states that exhibit probabilities greater than 0.05 at any time step during the evolution. 
\begin{rem}\label{remark-three}
The optimized pulses for the $\sqrt{\mathrm{SWAP}}$ gate
from Fig.~\ref{fig:higher_band_opt}(c)
result in a high state-evolution infidelity for error states with three and four atoms compared to states with one and two atoms. In particular, for shorter pulse durations, the atoms interact with each other and excite to the higher bands. However these error states 
can be suppressed with good initial state preparation, rendering them less significant for gate operations.
\end{rem}
If necessary, the optimization protocols can be extended to include infidelities from different atomic configurations in the cost function defined in Eq.~\eqref{gate_cost_function}, optimizing the full dynamics for more comprehensive experimental scenarios.

\begin{figure*}
\centering
\includegraphics{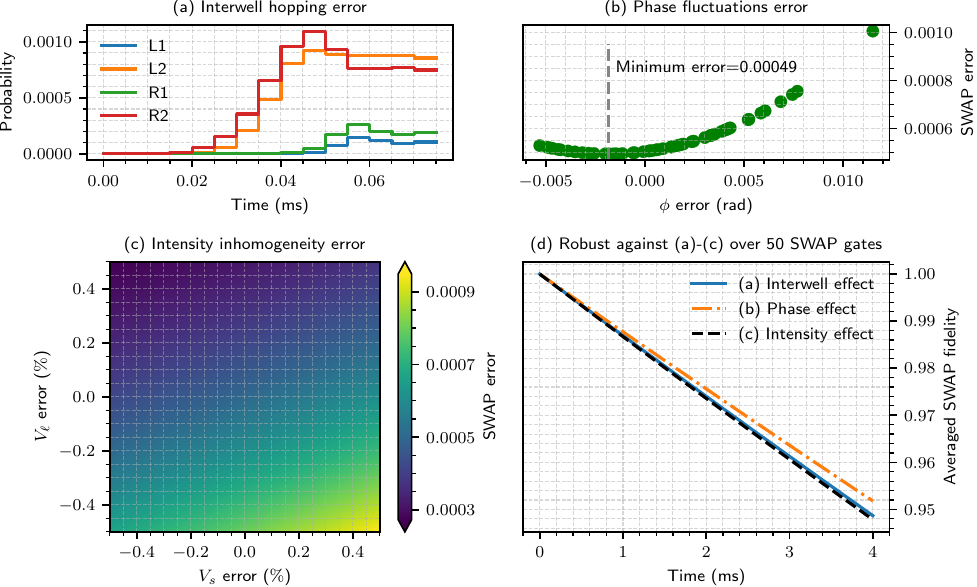}
\caption{Robustness of a SWAP gate against 
inter-well hopping, phase fluctuations, and intensity inhomogeneity
for an optimized control with a duration of $0.08$~ms:
(a)~Inter-well hopping for three double wells where $L1, L2, R1$ and $R2$ 
denote the probability of the atom being in the first and the second levels of the adjoining left and right double wells, respectively.
(b)~Phase fluctuations error by numerically introduced 
errors in the relative phase chosen from a Gaussian distribution with a standard deviation of $4.5$~mrad \cite{chalopin2024opticalsuperlatticeengineeringhubbard}.
(c)~Intensity inhomogeneity errors by numerically introduced
errors in the laser depths 
$V_s$ and $V_\ell$ chosen uniformly within $\pm 0.5 \%$ \cite{chalopin2024opticalsuperlatticeengineeringhubbard}.
(d) Exponential fit for the SWAP fidelity for each error source from (a)-(c) using 
$50$ SWAP gates. 
The intensity inhomogeneity is the most dominant source of error whereas phase fluctuations are the smallest.
For all errors, the exponential decay time of $\tau_d > 74$~ms (or $460$ gates) predicts a very long coherence time
for the chosen error strengths \cite{Impertro_2024, chalopin2024opticalsuperlatticeengineeringhubbard}.
\label{fig:robustness}}
\end{figure*}

\section{Robustness of the optimized control pulses\label{subsec:robustness}}
In this section, we assess the robustness of the optimal control pulses in the presence of different types of error sources \cite{Impertro_2024, chalopin2024opticalsuperlatticeengineeringhubbard}. 
The first error source involves potential tunneling from the target double well to neighboring wells. This means that during the gate operation, the atom has a finite probability of moving to the neighboring wells which reduces the gate fidelity. To test the robustness of the optimal control pulses against inter-well tunneling, we use a set of optimized $V_s$ and $V_\ell$ obtained from the SWAP optimization [see Sec.~\ref{higher_band_results}] of $0.08$~ms duration and $0.999$ fidelity as shown in Fig.~\ref{fig:higher_band_opt}(a). We simulate a system with three double wells where the middle double well is our target well and the left and right double wells act as the neighboring wells. We take only half of the adjacent left and right double wells resulting in eight possible states with two on the left double well, four on the middle double well, and two on the right double well for a single atom within the four-band model.  In Fig.~\ref{fig:robustness}(a), we present the time evolution of the states in the neighboring wells during the gate operation. Here, $L1$ and $L2$ represent the first and second levels of the left neighbor where $R1$ and $R2$ are the first and second levels of the right neighbor. When the atoms are in the first level, the overlap of the Wannier functions of different double wells is low, resulting in small hopping probabilities $L1$ and $R1$. The hopping probabilities $L2$ and $R2$ increase slightly when the atoms are in the second level due to the higher overlap of the Wannier functions, as shown in Fig.~\ref{fig:robustness}(a). The maximum probability of the atom tunneling to states outside the target double well at any time is negligible (${\approx }10^{-3}$) and the atom also tunnels back to the target double well since the probabilities in the neighboring wells decrease. This analysis considers the case where the neighboring double wells are initially empty. When the neighboring wells are occupied by atoms, the hopping probabilities change accordingly. For example, if there is an atom in the neighboring well with the same spin as the hopping atom, the hopping will be suppressed due to Pauli exclusion principle. However, if the atoms have opposite spins and the scattering length~$a$ is non-zero, the hopping will slightly increase due to a correction from the density-assisted hopping term, similar to Eq.~\eqref{eqn_for_delta_j}.

The second error comes from the phase instability of the lattice potential. In our simulations, we consider a symmetric double well with a relative phase of $\phi {=}0$, but in a real experiment, the relative phase fluctuations can lead to dephasing. Phase errors can generally be categorized into two types. The first type consists of time-independent or static phase fluctuations, which introduce a constant energy offset in the system. The second type consists of time-dependent phase fluctuations or drifts, which, if sufficiently strong, can lead to dephasing effects. In our work, we incorporate the static phase error into Eq.~\eqref{real_space_ham_1}, resulting in an energy offset between the left and right sites of the double well, thereby modifying the effective hopping amplitude. This alteration contributes to an increase in the gate error. When the gate is applied repeatedly over multiple cycles, this accumulated error leads to the damping of the Rabi oscillations, manifesting as a loss of coherence in the system's dynamics.
To understand the impact of these fluctuations, we sample 50 phase errors from a Gaussian distribution with a $4.5$~mrad standard deviation \cite{chalopin2024opticalsuperlatticeengineeringhubbard}.
We simulate our four-band model using these 50 phase errors with the same controls of $0.08$~ms duration.
As shown in  Fig.~\ref{fig:robustness}(b), the infidelity increases negligibly as $\phi$ changes, demonstrating the robustness of the controls against phase fluctuations.   

Lastly, we study the effect of the inhomogeneity in the laser intensities. These inhomogeneities lead to different coupling strengths across the lattice resulting in different gate errors. We test the robustness of the controls $V_s$ and $V_\ell$ against these fluctuations by uniformly selecting 400 pulses on a $20 {\times} 20$ grid
with extremal values given by the pairs $(V_s {\pm} e$, $V_\ell{\pm}e)$ and
an error of $e{\le} 0.5\%$ \cite{chalopin2024opticalsuperlatticeengineeringhubbard}. Figure~\ref{fig:robustness}(c) shows that the infidelity increases with increasing $V_s$ error as a result of decreasing tunneling strength $J$. For increasing $V_\ell$, $J$ increases and results in a decreasing infidelity.

To identify the dominant error sources among the three errors discussed, we run the simulations with each error source over 50 gate durations. This results in the damping of the Rabi oscillations and exponential decay of the SWAP fidelity over duration of $4$~ms as shown in Fig.~\ref{fig:robustness}(d). For phase and intensity errors, each data point is an average of the fidelity over 20 and 25 error pulses respectively. In agreement with the experimental observations in \cite{chalopin2024opticalsuperlatticeengineeringhubbard}, we see that inter-well hopping is one of the dominant sources of error whereas phase fluctuations cause the least damping. The larger effect of the intensity error can be explained by the fact that $V_s$ and $V_\ell$ are the controls for the optimization, so any variation in them has a significant effect on the fidelity. Note that while the damping of the Rabi oscillations appears similar to the effects caused by decoherence mechanisms such as dephasing and relaxation, our system remains closed even in the presence of additional noise sources. Therefore, in our case, what we refer to as dephasing corresponds to an accumulated coherent gate error rather than interaction with an external environment.
The fidelity after one gate duration is greater than $0.998$ for all of the error sources. The exponential decay time $\tau_d$ is greater than $74$~ms or $460$ Rabi oscillations for all errors. This is a predicted improvement of one order of magnitude compared to the experimental decay time $\tau_d {=}27$~ms  or $33$ oscillations \cite{chalopin2024opticalsuperlatticeengineeringhubbard}. This extended coherence time is primarily due to the faster gate duration of $0.08$~ms compared to the experimental gate duration of $0.4$~ms \cite{chalopin2024opticalsuperlatticeengineeringhubbard}.

\begin{figure}
\includegraphics{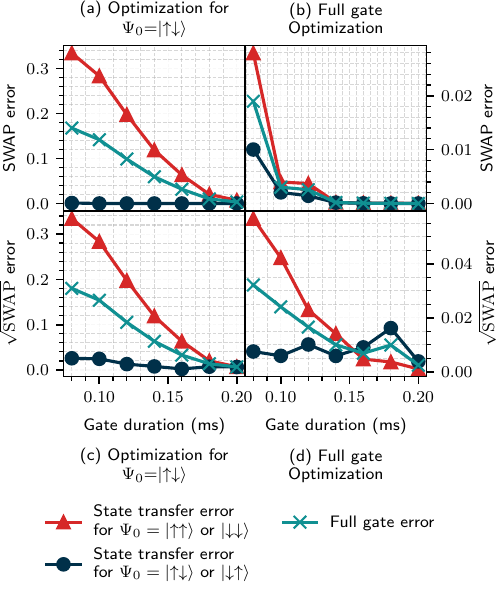}
\caption{Comparison of the optimization for one initial state $\Psi_0$ with the full gate optimization for different gate durations.
(a)~State evolution errors for SWAP and different initial states using optimized controls from Fig.~\ref{fig:higher_band_opt}(a). The initial states $\Psi_\text{ini}=\ket{\ua\da}$ and $\ket{\da\ua}$ have the same error due to the symmetry of the Hamiltonian and they have an error of less than 0.001 for gate durations larger than 0.08~ms [same as Fig.~\ref{fig:higher_band_opt}(a)]. The initial states $\Psi_\text{ini}=\ket{\ua\ua}$ and $\ket{\da\da}$ also follow the same dynamics but they have significantly higher errors for gate durations shorter than 0.20~ms. We show the full gate infidelity calculated with Eq.~\eqref{gate_cost_function}, which is larger than 0.01 for gate durations shorter than 0.20~ms. (b)~State evolution for SWAP and different initial states after the full gate optimization with the cost function of Eq.~\eqref{gate_cost_function}. Optimized controls from Fig.~\ref{fig:higher_band_opt}(a) are used as the initial guess for the full gate optimization. The state evolution error for $\Psi_\text{ini}=\ket{\ua\ua}$ and $\ket{\da\da}$ is significantly reduced compared to (a), and we have a full gate error of less than 0.005 for gate durations larger than 0.1~ms. (c)~Similar analysis is done for $\sqrt{\mathrm{SWAP}}$ and different gate durations and controls from Fig.~\ref{fig:higher_band_opt}(c).
(d)~The full gate error after further optimization
is in the range of $0.009{-}0.001$ for gate durations larger than 0.16~ms.
\label{fig:higher_band_full_gate}}
\end{figure}

\begin{res}\label{result-five}
The optimal control pulses are extremely robust against intensity inhomogeneity, phase fluctuations and inter-well hopping usually appearing in experiments. The inter-well hopping is one of the dominant sources of error whereas phase fluctuations cause the least damping which agrees with the experimental results \cite{chalopin2024opticalsuperlatticeengineeringhubbard}. The exponential decay time $\tau_d$ is greater than $460$ Rabi oscillations with gate duration of $0.08$~ms, which is almost an order of magnitude improvement compared to the experimental $\tau_d$ 
\cite{chalopin2024opticalsuperlatticeengineeringhubbard}.
\end{res}

\section{Full gate optimization with higher-band Fermi-Hubbard model\label{sec:full:gate}}
Recall for the two-band Fermi-Hubbard model from Sec.~\ref{numeric}
that the states $\ket{\ua\ua}$ and $\ket{\da\da}$ do not change under SWAP and the states $\ket{\ua\da}$ and $\ket{\da\ua}$ follow similar dynamics since the Hamiltonian is symmetric. 
Hence, the optimization for reaching the target state $\Psi_{\text{tar}}=\ket{\da\ua}$ also optimizes the full gate as shown in Fig.~\ref{two_band_opt}(a).

However, for the higher-band Fermi-Hubbard model, the states $\ket{\ua\ua}$ and $\ket{\da\da}$ can evolve to other states involving higher levels, e.g., for the four-band Hubbard model, $\ket{\ua\ua}$ or $\ket{\da\da}$ can have six possible states. An efficient SWAP or $\sqrt{\mathrm{SWAP}}$ should maximize the probability of the states $\ket{\ua\ua}$ and $\ket{\da\da}$ to stay in the lowest level at the end of the gate. To check the performance of our optimized controls from Fig.~\ref{fig:higher_band_opt}, we simulate the system with different initial states and calculate the infidelity of reaching the respective target states using Eq.~\eqref{cost_function} and the full gate error using Eq.~\eqref{gate_cost_function} with $N_i=4$. Figure~\ref{fig:higher_band_full_gate}(a) shows the state evolution errors for different initial states $\ket{\ua\ua},\ket{\da\da},\ket{\ua\da}$, and $\ket{\da\ua}$. The states $\ket{\ua\da}$ and $\ket{\da\ua}$ have the same errors as in Fig.~\ref{fig:higher_band_opt}(a) under SWAP gate. However, the states $\ket{\ua\ua}$ and $\ket{\da\da}$ have significantly higher errors for gate durations shorter than 0.20~ms since the atoms excite to the second level. This results in a full gate error larger than 0.01 for gate durations shorter than 0.20~ms.
We thus minimize the full gate error by minimizing the cost function defined in Eq.~\eqref{gate_cost_function} and using the controls from Fig.~\ref{fig:higher_band_opt}(a) as the initial guess. In Fig.~\ref{fig:higher_band_full_gate}(b), we show the state evolution errors after the full gate optimization. The errors with initial states $\ket{\ua\ua}$ and $\ket{\da\da}$ are significantly reduced, and we get a full gate error of less than 0.005 for gate durations larger than 0.10~ms.

We perform the same analysis for $\sqrt{\mathrm{SWAP}}$ and calculate the state evolution error for different initial states as shown in Fig.~\ref{fig:higher_band_full_gate}(c), using the optimized controls from Fig.~\ref{fig:higher_band_opt}(c). Similar to the SWAP gate, we see that the states $\ket{\ua\ua}$ and $\ket{\da\da}$ are exciting to higher levels, resulting in significant full gate error for gate durations less than 0.20~ms. After optimizing for all four initial states, we can decrease the state evolution errors and achieve a full gate error of less than 0.009 for gate durations larger than 0.16~ms [see Fig.~\ref{fig:higher_band_full_gate}(c)].
\begin{res}\label{full_gate_res}
With the full gate optimization, excitation to the higher bands is minimized for initial states $\ket{\ua\ua}$ and $\ket{\da\da}$, and the gate duration is found to be $0.10$~ms for SWAP with fidelity of $0.997$ and $0.16$~ms for $\sqrt{\mathrm{SWAP}}$ gate with fidelity of $0.993$.
\end{res}

\section{Effect of high interaction strength\label{sec:high:interaction}}
\begin{figure}
\includegraphics{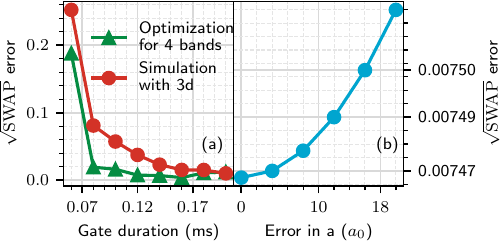}
    \caption{
    (a) Comparison of the $\sqrt{\mathrm{SWAP}}$ gate error between the one-dimensional four-band model and the full three-dimensional simulation for various gate durations. The increase in gate error for the three-dimensional case highlights the effect of excitations in the $y$ and $z$ directions.  
    (b) Robustness analysis of the $\sqrt{\mathrm{SWAP}}$ gate error against deviations in the scattering length. The gate error remains close to the ideal case for deviations $\Delta a$ ranging from $4a_0$ to $20a_0$, demonstrating negligible sensitivity of the optimized control pulses ($0.12$~ms gate duration) to scattering length deviations.
    }
\label{fig:y_z_comp}
\end{figure}
So far, we assume that there are no excitations in the $y$ and $z$ directions of the optical lattice. However, increasing the scattering length $a$ to high values can generate interactions in the $y$ and $z$ directions, resulting in excitations to higher levels. These excitations, if not suppressed, will result in a decrease of gate fidelities. The Hamiltonian for the three-dimensional system is written as 
\begin{align}
H_{3d}& = H_J^x \otimes I^y \otimes I^z + I^x \otimes H_E^y \otimes I^z \nonumber \\
& + I^x \otimes I^y \otimes H_E^z + U^x \otimes U^y \otimes U^z,
\end{align}  
where $H_J^x$ is the hopping plus the onsite energy in the $x$ direction, given by Eqs.~\eqref{hopping} and~\eqref{chemical_potential}, $H_E^y$ and $H_E^z$ are onsite energy terms in the $y$ and $z$ directions respectively, and $I$ is the identity matrix in different directions. Note that there is no hopping in the $y$ and $z$ directions. The interaction term $U^x$ is calculated from the one-dimensional form of Eq.~\eqref{eqn_for_U_general}, whereas $U^y$ and $U^z$ are calculated from an updated formula given by
($l \in \{y,z\}$)
\begin{equation}
U^{l} = \iint w_{m}^\dagger(l) \, w^\dagger_{n}(l) \, w_{o}(l) \, w_{p}(l) \, \mathrm{d}l.
\end{equation}   
%where $l_j =y_j$ if $l=y$,  $l_j =z_j$ if $l=z$ for $j \in \{1,2\}$.
We consider two energy levels, i.e., $m,n,o,p \in \{0,1\}$, resulting in four states for each of the $y$ and $z$ directions with two atoms. Combined with the four-band model in the $x$ direction, we have a total of 256 basis states for the three-dimensional simulation using $H_{3d}$.  
We simulate the three-dimensional model with optimized controls for different gate durations from Fig.~\ref{fig:higher_band_opt}(c) and calculate the $\sqrt{\mathrm{SWAP}}$ gate error. We compare the gate error from the three-dimensional simulation to the four-band one-dimensional model in Fig.~\ref{fig:y_z_comp}(a). For the gate duration of 0.12~ms, the infidelity of the $\sqrt{\mathrm{SWAP}}$ gate with the four-band model is 0.007, as shown in Fig.~\ref{fig:higher_band_opt}(c) and Fig.~\ref{fig:v_to_hubbard}(c). The infidelity with the three-dimensional model increases to 0.037 due to small excitations in the $y$ and $z$ directions. In principle, one can make a deeper lattice in the $y$ and $z$ directions compared to $45E_{ry}$ and $45E_{rz}$ used in our simulations to tightly squeeze the atoms. Additionally, we can also perform optimizations to suppress the excitations in these directions.

We optimize the time-independent scattering length $a$ to arbitrary values with respect to an upper bound. However, tuning the magnetic field to achieve the exact $a$ value is difficult. Therefore, we analyze the robustness of the $\sqrt{\mathrm{SWAP}}$ gate against deviations in the scattering length. For this, we simulate the system with the four-band Fermi-Hubbard model using the optimized controls from Fig.~\ref{fig:v_to_hubbard}(a) $0.12$~ms gate duration, and $a = 1995.22 \times a_0$. We take different deviation $\Delta a$ values from $4a_0$ to $20a_0$, resulting in an effective scattering length of $a+\Delta a$. As shown in Fig.~\ref{fig:y_z_comp}(b), the increase in gate error is negligible for this range of deviation, showing the robustness of our optimized pulses.

\section{Conclusion}\label{Conclusion}
Using open-loop quantum optimal control, we design pulses for the SWAP and $\sqrt{\mathrm{SWAP}}$ gates for fermionic atoms trapped in a superlattice reaching fidelities in the range of $10^{-3}$ for realistic experimental parameters.
We use Fermi-Hubbard models for the optimization, which widely describe fermionic atoms trapped in optical lattices or superlattices. 
In this work, we treat the gate optimization as a state-to-state transfer 
for the SWAP and $\sqrt{\mathrm{SWAP}}$ gates using the two-band Fermi-Hubbard model. So,
optimizing a SWAP gate corresponds to optimizing the transfer of the 
state $\ket{\ua\da}$ to $\ket{\da\ua}$. A $\sqrt{\mathrm{SWAP}}$ gate optimization corresponds
to optimization of transfer of the state $\ket{\ua\da}$ to $[{(1{+}i)}\ket{\ua\da} {- {(1{-}i)}}\ket{\da\ua}]/2$.
First, using the two-band Fermi-Hubbard model, we find the time-optimal control for the SWAP gate analytically and calculate the quantum speed limit in the presence of bounds on the control. We match the analytical study with the numerical optimization of the SWAP gate and show a numerical quantum speed limit closely matching with the analytical one with feasibility constraints. 
We also calculate the full gate fidelities, which is the summation of state transfer fidelities for all computational states, and show that gate fidelities are equivalent to the optimized state transfer fidelities.

Next, we show the limitations of the two-band Fermi-Hubbard model in the case of fast gates, where it is not sufficient to provide the full dynamics of the double well system. We describe higher-band Fermi-Hubbard models and update our time-evolution method to account for the non-adiabatic change of the Hamiltonian. We detail a formula to calculate the approximate analytical gradient of the cost function and use a combination of this and the finite-difference method in the optimization. We find that optimization with four bands is sufficient for the SWAP gate, whereas the performance of the $\sqrt{\mathrm{SWAP}}$ gate is further improved by a six-band optimization. Our numerical simulations demonstrate that high-fidelity SWAP and $\sqrt{\mathrm{SWAP}}$ gates can be realized with significantly reduced control durations using realistic experimental parameters. We achieve a SWAP between the states $\ket{\ua\da}$ and $\ket{\da\ua}$ in $0.08$~ms with fidelity of $0.999$ and a $\sqrt{\mathrm{SWAP}}$ pulse preparing the entangled state $[{(1{+}i)}\ket{\ua\da} {- {(1{-}i)}}\ket{\da\ua}]/2$ in $0.12$~ms with fidelity of $0.995$, representing a five-fold improvement over experimental state transfer times \cite{Impertro_2024, chalopin2024opticalsuperlatticeengineeringhubbard}. These findings emphasize the potential of higher-band Fermi-Hubbard models for accelerating and optimizing quantum gates.

We test our optimal control pulses for $\sqrt{\mathrm{SWAP}}$ gate on the different atom configurations in the double well. The three-atom and four-atom error states observe significant errors, especially for short gate durations, while these errors can be optimized if necessary.
We check the robustness of the optimal control pulses against inter-well hopping, phase fluctuations, and intensity inhomogeneity.
We show that the optimal pulses are extremely robust against all errors with inhomogeneous intensity and inter-well hopping being the dominant ones.  
Furthermore, by employing the full gate optimization, we minimize excitations to higher bands for initial states $\ket{\uparrow\uparrow}$ and $\ket{\downarrow\downarrow}$, achieving a SWAP gate 
fidelity of $0.997$ in $0.10$~ms and a $\sqrt{\mathrm{SWAP}}$ gate fidelity of $0.993$  in $0.16$~ms.
We also explain the different gradient methods used in the paper and compare their performance based on the number of function evaluations and the optimization duration. 
Our work therefore performs a detailed study of fermionic atoms trapped in a superlattice system and designs faster and more efficient gates. One can readily adapt the modeling and optimization scheme of this work for bosons as well by employing the Bose-Hubbard model. Unlike fermions, bosons differ in their interaction energy, which permits double occupancy by atoms with the same spin states. Furthermore, the creation and annihilation operators for bosons obey the standard commutation relations, in contrast to the fermionic anti-commutation relations. 

Our optimization model is inspired by experiments and incorporates realistic parameters \cite{chalopin2024opticalsuperlatticeengineeringhubbard}. The experiments performed in \cite{chalopin2024opticalsuperlatticeengineeringhubbard} use phase modulation to perform the SWAP operation, whereas our optimization uses the lattice depths $V_s$ and $V_l$ as the control~\cite{Impertro_2024}. This gives the advantage of minimizing the effect of phase noise, which is critical in optical lattice experiments. 
The pulses generated from optimizing the SWAP and $\sqrt{\mathrm{SWAP}}$ gates are both reasonably smooth and experimentally implementable. Thus applying these optimized controls into real experiments will enable higher gate fidelities. For more realistic pulses, we can include the transfer function information in the optimization \cite{PhysRevApplied.19.064067}. The optimized pulses must work with minimal excitation in the $y$ and $z$ directions, which can be achieved with deeper $y$ and $z$ lattice depths. Experiments utilizing optimal control will not only enhance gate performance, but they will enable us to assess and refine our existing models, such as by accounting for heating effects.
Moreover, pulses optimized by open-loop approaches are a good starting point for further feedback-based optimizations directly applied in experiments. This iterative approach will pave the way for robust and efficient quantum gates,
which are crucial for quantum simulation and computation.

\begin{acknowledgments}
We would like to thank Petar Bojovic, Robin Groth, Titus Franz, Philipp Preiss, and Timon Hilker for insightful discussions about the fermionic system as well as for providing detailed experimental
parameters.
J.S. would like to thank Ashutosh Mishra and Kunal Vyas for illuminating conversations on Fermi-Hubbard model.
We acknowledge support from the German Federal Ministry of Education and Research through
the funding program quantum technologies—from basic research to market under the projects FermiQP, \href{https://www.quantentechnologien.de/forschung/foerderung/quantenprozessoren-und-technologien-fuer-quantencomputer/fermiqp.html}{13N15891}, and MUNIQC-Atoms, \href{https://www.quantentechnologien.de/forschung/foerderung/quantencomputer-demonstrationsaufbauten/muniqc-atoms.html}{13N16073}, as well as via the Helmholtz Validation Fund project Qruise (HVF-00096).
\end{acknowledgments}
\appendix
\section{Analytical SWAP optimization}\label{app:analytic}
In this appendix, we find the time-optimal control $J(t)$ and the analytical quantum speed limit for SWAP gate. We solve the Euler-Lagrange equations for the two-band Fermi-Hubbard model described in Sec.~\ref{analytical}. 
We show in Sec.~\ref{objective} that for the two-band Fermi-Hubbard model, optimizing the transfer from state $\ket{\ua\da}$ to $\ket{\da\ua}$ is sufficient to optimize the SWAP gate.
As described in Sec.~\ref{analytical},
the time-evolution with $\tilde{J}(t){=}{-}\sqrt{2}J(t)$ and $U{=}0$ is
\begin{equation}\label{eqn_for_motion_mat:app}
i\begin{pmatrix}
\dot{x}_1\\
\dot{x}_2\\
\dot{x}_3
\end{pmatrix}
=
\begin{bmatrix}
0 & 0 & \tilde{J}(t)  \\
0 & 0 & \tilde{J}(t)  \\
\tilde{J}(t)& \tilde{J}(t) & 0 \\
\end{bmatrix}
\begin{pmatrix}
x_1\\
x_2\\
x_3
\end{pmatrix}
\end{equation}
where $x_1, x_2$, and $x_3$ correspond to complex coefficients of the states $\ket{\ua\da}$,  $\ket{\da\ua}$, and $(\ket{D0} {+}\ket{0D})/\sqrt{2}$ respectively. Thus, optimizing the SWAP gate reduces to finding $\tilde{J}(t)$ for evolving the system
from $(1,0,0)^T$ to $(0,1,0)^T$, or at least up to a phase factor as we see below.

With $x_1=r_1{+}ir_4, x_2=r_2{+}ir_5$, $x_3=r_6{+}ir_3$, and real $r_j=r_j(t)$ and $\tilde{J}(t)$, we obtain
\begin{align*}
&\dot{r}_1=\tilde{J}(t)r_3,\;
\dot{r}_2=\tilde{J}(t)r_3,\;
\dot{r}_3=-\tilde{J}(t)r_1{-}\tilde{J}(t)r_2 \;\text{ and} \\
&\dot{r}_4=-\tilde{J}(t)r_6,\;
\dot{r}_5=-\tilde{J}(t)r_6,\;
\dot{r}_6=\tilde{J}(t)r_4+\tilde{J}(t)r_5. \nonumber
\end{align*}
So the variables $r_1$, $r_2$, and $r_3$ are decoupled from $r_4$, $r_5$, and $r_6$. We obtain two independent
three-dimensional subsystems and we work with the first subsystem of them \cite{Boscain-Rudolf}.
The ordinary differential equations are given by
\begin{equation}
\begin{pmatrix}
\dot{r}_1\\
\dot{r}_2\\
\dot{r}_3
\end{pmatrix}
=
\begin{bmatrix}
0 & 0 & \tilde{J}(t)  \\
0 & 0 & \tilde{J}(t)  \\
-\tilde{J}(t)& -\tilde{J}(t) & 0 \\
\end{bmatrix}
\begin{pmatrix}
r_1\\
r_2\\
r_3
\end{pmatrix}.
\label{eqn_for_motion_mat_real}
\end{equation}
Clearly, $r_1^2+r_2^2+r_3^2=1$, provided we start in the subspace $r_1,r_2$, and $r_3$. From Eq.~\eqref{eqn_for_motion_mat_real}, 
\begin{equation}
\dot{r}_1=\dot{r}_2 \;\text{ implies }\;
r_1=r_2+C
\label{eq:constant:C}
\end{equation}
for a suitable real constant $C$.
Minimizing the transfer time $T$ is equivalent \cite{Boscain-Rudolf} to minimizing the functional 
\begin{equation*}
    \mathcal{E}= \int_{0}^{T} \tilde{J}^2(t)\, dt = \int_{0}^{T}  \mathcal{L}\, dt,
\end{equation*}
and $\tilde{J}(t)=-\sqrt{2} J(t)$ needs to be bounded for the optimal solution to be well defined.
Applying Eq.~\eqref{eqn_for_motion_mat_real},
the Lagrangian of the system is given by
\begin{equation*}
\mathcal{L}= \tilde{J}^2(t) = \frac{{\dot{r}_1^2}}{{r^2_3}}=\frac{{\dot{r}_1^2}}{{1{-}r_1^2{-}r_2^2}}
=\frac{{\dot{r}_1^2}}{{1{-}r_1^2{-}(r_1{-}C)^2}}.
\end{equation*}
The upper bound on the control can be re-normalized by the maximum amplitude
that is possible in an experiment. 
We use the Euler-Lagrange equations
\begin{equation}\label{euler_lag}
    \frac{d}{dt}\left[\frac{\partial \mathcal{L}}{\partial \dot{r}_1}\right] =\frac{\partial \mathcal{L}}{\partial {r_1}}
\end{equation}
to find the optimal solution.
Computing the left hand side of Eq.~\eqref{euler_lag}, we have
\begin{equation*}
\frac{\partial \mathcal{L}}{\partial \dot{r}_1}=\frac{2\dot{r}_1}{r_3^2}
\;\text{ which implies }\;
\frac{d}{dt}\left[\frac{\partial \mathcal{L}}{\partial \dot{r}_1}\right]= 2\frac{d}{dt} \left[\frac{\dot{r}_1}{r_3^2}\right]
\end{equation*}
and the corresponding right hand side is given by
\begin{align*}
& \frac{\partial \mathcal{L}}{\partial {r_1}} = \frac{\partial }{\partial {r_1}} \Big[\frac{{\dot{r}_1^2}}{{1{-}r_1^2{-}r_2^2}}\Big]\\
&=-\frac{\dot{r}_1^2}{({1{-}r_1^2{-}r_2^2})^2}\frac{\partial }{\partial {r_1}} [1{-}r_1^2{-}(r_1{-}C)^2]
=\frac{\dot{r}_1^2}{r_3^4}(4r_1{-}2C).
\end{align*}
Therefore, Eq.~\eqref{euler_lag} simplifies to
\begin{align}\label{intermediate_eqn}
\frac{d}{dt} \left[\frac{\dot{r}_1}{r_3^2}\right]&=\frac{\dot{r}_1^2}{r_3^4}(2r_1{-}C).
\end{align}
We separately compute
\begin{align*}
\frac{d}{dt} \left[\frac{\dot{r}_1}{r_3^2}\right]&=\frac{1}{r_3} \frac{d}{dt} \left[\frac{\dot{r}_1}{r_3}\right]+\frac{\dot{r}_1}{r_3}\frac{d}{dt}\left[\frac{1}{r_3}\right]\\
&=\frac{1}{r_3} \frac{d}{dt} \left[\frac{\dot{r}_1}{r_3}\right]-\frac{\dot{r}_1}{r_3^3}\big[{-}\tilde{J}(t)r_1{-}\tilde{J}(t)r_2\big]\\
&=\frac{1}{r_3} \frac{d}{dt} \left[\frac{\dot{r}_1}{r_3}\right]-\frac{\dot{r}_1}{r_3^3}\big[{-}\frac{\dot{r}_1}{r_3}r_1{-}\frac{\dot{r}_1}{r_3}(r_1{-}C)\big]\\
&=\frac{1}{r_3} \frac{d}{dt} \left[\frac{\dot{r}_1}{r_3}\right]+\frac{\dot{r}^2_1}{r_3^4}(2r_1{-}C)
\end{align*}
and we substitute this back into Eq.~\eqref{intermediate_eqn} and obtain
\begin{equation*}
 \frac{d}{dt} \left[\frac{\dot{r}_1}{r_3}\right]=0=\frac{d\tilde{J}(t)}{dt}.
\end{equation*}
This finally implies that $\tilde{J}(t)$ and $J(t)$ are constant.
We substitute $\tilde{J}(t)={-}\sqrt{2}A $ in Eq.~\eqref{eqn_for_motion_mat:app} where $A$ is a suitable constant.
We now consider the initial conditions $x_1(0)=1$, $x_2(0)=0$, and $x_3(0)=0$.
Note that this implies $C=1$ in Eq.~\eqref{eq:constant:C}.
For these initial conditions,
we directly solve Eq.~\eqref{eqn_for_motion_mat:app} (which has now only constant coefficients)
and obtain
\begin{equation}\label{analytic_soln}
x_1=\cos^2(At),\,
x_2=-\sin^2(At),\,
x_3={i\sin(2At)}/{\sqrt{2}}.
\end{equation}
For the target state $(0,-1,0)^T$,
we get $AT={\pi}/{2}{+}n\pi$. Hence, the fastest transfer from $(1,0,0)^T$ to 
$(0,-1,0)^T$ can be attained in a time of $T={\pi}/(2A)$. This defines our quantum speed limit for the SWAP gate and is given by the constant hopping parameter $J(t)=A$. 
Clearly, $J(t)=A$ needs to be bounded by a maximal allowed $J_{\text{max}}$, i.e.\ $J(t) \leq J_\text{max}$.
The value of $J_\text{max}$ depends on the experimental setup and we set $J_\text{max}=34.03$ kHz for the two-band
numerical simulations in Sec.~\ref{numeric}.
That means the quantum speed limit for the two-band SWAP gate is $T={\pi}/{(2J_\text{max})}$. 
We obtain the control $J(t)=A=34.03$ kHz and the corresponding quantum speed limit is $T=0.046$~ms.

\section{Approximate analytical gradient\label{approximate-analytic-gradient}}
As explained in Sec.~\ref{optimization with higher band}, we can not use the fast spline-fit method (described in Sec.~\ref{numeric} and Appendix~\ref{sec:gradient:methods}) for calculating the gradients of the cost function $C$ for optimizations using higher-band models. In this appendix, we derive an approximate analytical formula for the gradients and use it in the optimizations performed in Sec.~\ref{higher_band_results}.
Our primary focus is to find an analytical formula for calculating
${d C}/{dV_{k}}$ where $V_k \in \{V_s, V_\ell\}$. 
The cost function $C$ for the target state $\Psi_{\text{tar}}$ and the modified time evolution of Eq.~\eqref{time_evol_proj} is 
\begin{equation*}\label{cost_func_higher}
C{=}1{-} \abst{\langle \Psi_{\text{tar}} \vert\, \mathcal{U}_{N_T}\cdot\cdot P(w_{t+1},\!w_t)\mathcal{U}_{t}P(w_{t},\!w_{t-1})\cdot\cdot\, \mathcal{U}_{1} \vert \Psi_\text{ini}\rangle}^2. 
\end{equation*}
Using the product rule, one infers that ${d C}/{dV_{k}}$ has three terms proportional to ${\partial P(w_{t+1},w_t)}/{\partial V_{k}}$,  ${\partial \mathcal{U}_{t}}/{\partial V_{k}}$, and  ${\partial P(w_{t},w_{t-1})}/{\partial V_{k}}$.
We approximate the gradient of the unitary evolution operator $ \mathcal{U}_{t}$ as
\begin{equation}\label{du_dv}
\frac{\partial e^{-iH_t\partial t}}{\partial V_{k}}= -iH'_t\partial t e^{-iH_t\partial t},
\end{equation}
where $H_t$ is the higher-band Fermi-Hubbard Hamiltonian \eqref{fermi_hubbard_higher} at time step $t$. To calculate $H'_t$ from \eqref{fermi_hubbard_higher}, we need to calculate ${\partial J_p}/{\partial V_{k}}$ and ${\partial \epsilon_{pm}}/{\partial V_{k}}$; all other terms
proportional to $a$ are zero.  
The hopping parameter $J_p$ and the onsite energy $\epsilon_{p}$ is calculated from Eqs.~\eqref{v_to_j_general} and \eqref{onsite energy} which can also be written as 
\begin{equation*}
J_p=\frac{{E_{2p+1}-E_{2p}}}{2} \; ; \; \epsilon_p=\frac{{E_{2p+1}+E_{2p}}}{2},
\end{equation*} 
where $E_j$ is the energy of the $j${th} band of the double well. 
Therefore we can calculate ${\partial J_p}/{\partial V_{k}}$ and ${\partial \epsilon_{pm}}/{\partial V_{k}}$ as

\begin{equation}\label{dj_dv}
\frac{\partial J_p}{\partial V_{k}}= \frac{1}{2}({\frac{\partial E_{2p+1}}{\partial V_{k}} {-}\frac{\partial E_{2p}}{\partial V_{k}})},\,
\frac{\partial \epsilon_p}{\partial V_{k}}=\frac{1}{2}({\frac{\partial E_{2p+1}}{\partial V_{k}} {+}\frac{\partial E_{2p}}{\partial V_{k}})}.
 \end{equation}
Moreover, ${\partial E_i}/{\partial V_{k}}$ can be calculated from
\begin{equation}
\frac{\partial E_i}{\partial V_{k}}= \sum_{j}v_{ij}^{{\dagger}}\frac{\partial \tilde{H}_1(q)}{\partial V_{k}}v_{ij},       
\label{dE_dv}  
\end{equation}
where $\tilde{H}_1(q)v_{ij}=E_{ij}v_{ij}$ and $v_{ij}^{{\dagger}}v_{ij}=1$ \cite{Wilkinson1965}. Here, $\tilde{H}_1(q)$ denotes the Fourier transform of the Hamiltonian in Eq.~\eqref{potential_2} and $v_{ij}$ are the eigenstates of $\tilde{H}_1(q)$  [see Sec.~\ref{Two-band-Fermi-Hubbard} and Eq.~\eqref{fourier_ham}]. Thus we can calculate ${\partial \mathcal{U}_t}/{\partial V_{k}}$ using Eq.~\eqref{du_dv}-\eqref{dE_dv} and eventually the final gradient $ {dC}/{dV_{k}}$.
Next, the basis transformation operator $P(w_{t+1},w_t)$ depends on the Wannier function $w_t$ so that
\begin{equation*}
\frac{\partial P(w_{t+1},w_t)}{\partial V_{k}}=P(w_{t+1},\frac{\partial w_{t}}{\partial V_{k}}). 
\end{equation*}
The gradient of the Wannier functions is calculated via \cite{Wilkinson1965}
\begin{equation*}
\frac{\partial w_i}{\partial V_{k}} = -\sum_{j}\, [\tilde{H}_1(q){-}E_{ij}\mathcal{I}]^{{+}}\;
\Big[\frac{\partial \tilde{H}_1(q)}{\partial V_{k}}{-}\frac{\partial E_{ij}}{\partial V_{k}}\Big]\, v_{ij},   
\end{equation*}
where $\mathcal{I}$ is the identity operator and $[\tilde{H}_1(q){-}E_{ij}\mathcal{I}]^{{+}}$ is the Moore-Penrose inverse of $[\tilde{H}_1(q){-}E_{ij}\mathcal{I}]$. The contribution of ${\partial P(w_{t+1},w_t)}/{\partial V_{k}}$ is negligible in the final gradient ${dC}/{dV_{k}}$. So, we set this term to zero, speeding up the calculations by avoiding several matrix multiplications. 
Similarly, we ignore the term ${\partial P(w_{t},w_{t-1})}/{\partial V_{k}}$.  
This approximate analytical gradient computation is significantly faster than the finite-difference method, as shown in Appendix~\ref{sec:gradient:methods}. However, it becomes difficult to optimize pulses further when we are getting close to the minimum since the approximate analytical method does not provide an exact gradient.
Therefore, we reach the minimum SWAP- and $\sqrt{\mathrm{SWAP}}$-gate error by combining
the approximate analytical and finite-difference methods.

\section{Comparison of different methods to compute gradients\label{sec:gradient:methods}}
Throughout the paper, we have used gradient-based methods for optimizing the lattice depths $V_{s}$ and $V_\ell$ and the scattering length $a$. We optimize $V_s$ and $V_\ell$ for the SWAP gate, and $V_s$, $V_\ell$, and $a$ for the $\sqrt{\mathrm{SWAP}}$ gate using GRAPE-like algorithms. The gradient calculation for the optimization of $a$ is trivial and finite differences work efficiently since we need to optimize only one parameter. However, $V_s$ and $V_\ell$ are time-dependent and piecewise-constant controls, so we need more sophisticated and faster ways of calculating the gradient.  
One straightforward way for calculating gradients is to use the finite-difference method to calculate ${d C}/{d V_{k}}$ for $k\in\{s,\ell\}$ at every time step. We use the efficient built-in finite-difference implementation from Scipy \cite{2020SciPy-NMeth} for the comparison in this section. 
We can also calculate the gradient analytically using the approximation discussed in Sec.~\ref{approximate-analytic-gradient}.
For the two-band optimization performed in  
Sec.~\ref{numeric}, we use the spline-fit method to calculate the gradients. For the spline-fit method, we first store a data set of triples $(V_s, V_\ell, J)$. From the stored data set, we fit a spline function $J{=}S(V_s, V_\ell)$ over the grid of pairs $(V_s, V_\ell)$. We use the SciPy package \cite{2020SciPy-NMeth} for the spline-fit and this function also provides us with the gradient of the estimated $J$ at any pair $(V_s, V_\ell)$. We use these gradients in the GRAPE algorithm to run the full optimization (refer to Sec.~\ref{numeric} for details). One can also use automatic differentiation for gradient calculation \cite{10.5555/3122009.3242010}, but it can be slow in the presence of multiple matrix diagonalizations for calculating the Wannier states. Hence, we do not analyze the performance of automatic differentiation here. Two parameters are used to test the efficiency of different gradient methods. First, we check the number of cost function evaluations 
in the optimization for a particular gradient method. Secondly, the total optimization run times 
are compared for different gradients.

\begin{figure}
 \includegraphics{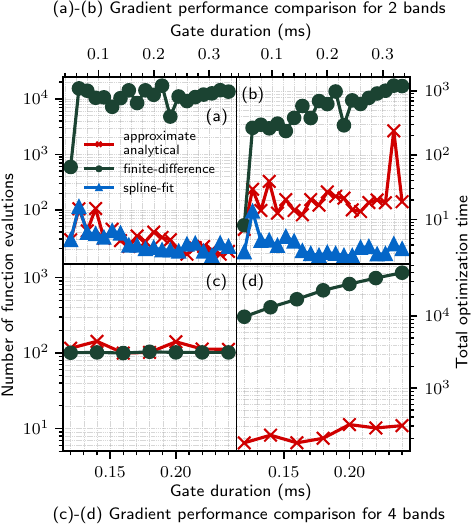}
\caption{Performance of the approximate analytical, the finite-difference, and the spline method
for computing the gradient
for two-band (a)-(b) and four-band (c)-(d) Fermi-Hubbard optimizations:
(a)-(b)~Overall spline method performs best both in the number of evaluations and the total optimization time
and remains mostly constant regardless of the gate duration,
while the finite-difference method preforms worst and its total optimization time increases with the gate duration.
(c)-(d)~The number of function evaluations for the approximate analytical gradient computation
is comparable with the finite-difference method, 
which however performs much worse for the total optimization time. The spline method
is not available for four bands.
\label{fig:grad_comp}}
\end{figure}

We compare the gradient methods for the two-band model at varying gate durations. Figure~\ref{fig:grad_comp}(a) shows that the finite-difference method results in the highest number of function evaluations for every gate duration, whereas the approximate analytical and spline-fit methods are comparable. Function evaluations 
decrease for longer gate durations due to lower initial gate error and easier optimization tasks for longer durations. Figure~\ref{fig:grad_comp}(b) provides insights into the performance of different gradient methods. The finite difference method is the slowest to reach the minima and optimization times increase linearly with the gate duration. The approximate analytical and spline-fit method
result in run times that are relatively independent of the gate duration. But the approximate analytical gradient needs a longer time compared to the spline fit. As the spline fit calculates $J$ and ${d C}/{d V_{k}}$ from existing fitted functions, it avoids several matrix diagonalizations and multiplications needed for calculating the Wannier states, which render the approximate analytical gradient computationally more
expensive.

A similar comparison is done for the four-band model where we compare the performance of the approximate analytical gradient with the
finite-difference gradient. Note, that since we have to calculate basis transformation operators at every time step, we cannot use the spline-fit method for higher-band optimization. We set an upper bound on the number of function evaluations for both the finite-difference and the approximate analytical method. As we can see in Fig.~\ref{fig:grad_comp}(c) both methods have maximal function evaluations for all gate durations. However, finite differences lead to longer optimization times when compared to the approximate analytic method as shown in Fig.~\ref{fig:grad_comp}(d). This suggests that the spline-fit method performs significantly better than the approximate analytical and finite-difference methods, but for higher bands we cannot use the spline-fit method. In that case, the use of approximate analytical gradients is computationally effective. However, since the approximate analytical method does not give an exact gradient, we use a combination of the approximate analytical and finite-difference methods to minimize the cost function in Sec.~\ref{higher_band_results}.
Hence, we use multiple methods for gradient calculation depending on the complexity of the system. Different methods have different trade-offs and combining them enables fast and efficient optimizations.

\begin{table}[t]
\centering
\begin{tabular}{c c c c c}
\hline\hline
\\[-2.5mm]
$\uparrow$ & \hspace{0.5cm} $\downarrow$ & \hspace{0.5cm} $I_\uparrow$ & \hspace{0.5cm} $I_\downarrow$ & \hspace{0.5cm} $I=I_{\uparrow}\binom{S}{N_\da}+I_{\downarrow}$ \\[1.5mm]
\hline
\\[-2.5mm]
0001 & \hspace{0.5cm} 0001 & \hspace{0.5cm} 0 & \hspace{0.5cm} 0 & \hspace{0.5cm} 0 \\
0001 & \hspace{0.5cm} 0010 & \hspace{0.5cm} 0 & \hspace{0.5cm} 1 & \hspace{0.5cm} 1 \\
: & \hspace{0.5cm} : & \hspace{0.5cm} : & \hspace{0.5cm} : & \hspace{0.5cm} : \\
0010 & \hspace{0.5cm} 0001 & \hspace{0.5cm} 1 & \hspace{0.5cm} 0 & \hspace{0.5cm} 4 \\
: & \hspace{0.5cm} : & \hspace{0.5cm} : & \hspace{0.5cm} : & \hspace{0.5cm} : \\
1000 & \hspace{0.5cm} 1000 & \hspace{0.5cm} 3 & \hspace{0.5cm} 3 & \hspace{0.5cm} 15 \\[0.5mm]
\hline\hline
\end{tabular}
\caption{A scheme explaining the calculation and labeling of the computational states for $N_\uparrow=1$ and $N_\downarrow=1$. In the first column, the total number of bits is the number of sites and the number of ones in each bit configuration represents the number of spins up.
The same applies to the second column, where the spins up are replaced with the spins down.
 The states are ordered first with spin-up (first and third column) and then with spin-down (second and fourth column). The final label
 (fifth column) is computed from Eq.~\eqref{basis_label}.}
\label{table:table_state}
\end{table}

\section{Assignment of computational basis\label{app:basis_labeling}}
In this appendix, we describe the method used for finding computational basis states and labeling them for different number of atoms in a double well. We use this labeling method in Sec.~\ref{three_four_atom} to simulate the dynamics of different atom configurations.
Suppose we have $M$ levels and $N$ atoms in the double well. Now since each level consists of the left and right sides of the double well, we have a total of $S{=}2{\times} M$ sites. Now suppose, we have $N_{\ua}$ atoms with spin up and $N_{\da}$ with spin down, i.e.\ $N{=}N_{\ua}{+}N_{\da}$. We can arrange $N_{\ua}$ spin-up identical fermionic atoms in $\binom{S}{N_\ua}{=}{S!}/{N_{\ua}!(S{-}N_{\ua})!}$ ways in $S$ sites. For each configuration of the spin-up atoms, we can arrange the spin-down atoms in $\binom{S}{N_\da}{=}{S!}/{N_{\da}!(S{-}N_{\da})!}$ ways. The total number of possibilities we can arrange $N{=}N_{\ua}{+}N_{\da}$ atoms in a double well is the number of computational states and is given by $N_B=\binom{S}{N_\ua} \times \binom{S}{N_\da}$.
As an example, Table~\ref{table:table_state} shows the possible computational basis states for two atoms with one spin-up and one spin-down used in the four-band Fermi-Hubbard model in Sec.~\ref{optimization with higher band}. 
Here $M{=}2, S{=}4, N{=}2, N_{\ua}{=}1,$ and $N_{\da}{=}1$.
So the total number of computational states is $16$. We represent the spin-up and spin-down state by $S$ bits and label our computational state using the following formula
\begin{equation}\label{basis_label}
I=I_{\uparrow}\binom{S}{N_\da}+I_{\downarrow},
\end{equation}
where these values
observe $I_\ua \in \{0,\ldots,\binom{S}{N_\ua}{-}1\}$,  $I_\da \in \{0,\ldots, \binom{S}{N_\da}{-}1\}$,
and $I \in \{0,\ldots,N_B{-}1\}$.

%\bibliography{Paper}
%apsrev4-2.bst 2019-01-14 (MD) hand-edited version of apsrev4-1.bst
%Control: key (0)
%Control: author (8) initials jnrlst
%Control: editor formatted (1) identically to author
%Control: production of article title (-1) disabled
%Control: page (0) single
%Control: year (1) truncated
%Control: production of eprint (0) enabled
%

\end{document}